\documentclass[longbibliography,reprint,2,amsmath,amssymb,aps]{revtex4-1}

\usepackage{graphicx}
\usepackage{dcolumn}
\usepackage{bm}
\usepackage{natbib}
\usepackage[table]{xcolor}
\usepackage{tabularx}
\usepackage{array}
\usepackage{colortbl}
\usepackage{appendix}
\usepackage{amsmath}
\DeclareMathOperator{\sinc}{sinc}

\begin{document}
\preprint{APS/123-QE}

\title{Harnessing Hybrid Frequency-Entangled Qudits through Quantum Interference}

\author{Sheng-Hung Wang $^{1,3}$, Po-Han Chen $^{1,3}$, Cheng-Yu Yang$^{2}$
Yen-Hung Chen $^{2,3,\dagger}$, Pin-Ju Tsai $^{2,3}$}
\email{tpinju@ncu.edu.tw}
\email{$^{\dagger}$yhchen@dop.ncu.edu.tw}

\affiliation{$^{1}$Department of Physics, National Central University, Taoyuan City 320317, Taiwan
}
\affiliation{$^{2}$Department of Optics and Photonics, National Central University, Taoyuan City 320317, Taiwan
}
\affiliation{$^{3}$Quantum Technology Center, National Central University, Taoyuan City 320317, Taiwan
}

\date{\today}
\begin{abstract}


High-dimensional (HD) quantum entanglement expands the Hilbert space, offering a robust framework for quantum information processing with enhanced capacity and error resilience. In this work, we present a novel HD frequency-domain entangled state, the hybrid frequency-entangled qudit (HFEQ), generated via Hong-Ou-Mandel (HOM) interference, exhibiting both discrete-variable (DV) and continuous-variable (CV) characteristics in the frequency domain. By tuning HOM interference, we generate and control HFEQs with dimensions $D=5,7,9,11$, confirming their DV nature. Franson interferometry confirms the global frequency correlations with visibility exceeding 98\% and verifies the CV entanglement within individual frequency modes with visibility greater than 95\%. Our findings provide deeper insight into the physical nature of frequency-entangled qudits generated by quantum interference and introduce a novel resource for HD time-frequency quantum information processing.

\end{abstract}
\maketitle
\section{Introduction}

High-dimensional (HD) quantum information processing (QIP) based on $D$-dimensional qudits plays a crucial role in the development of large-scale quantum computing and quantum communication with high information capacity and error tolerance \cite{barreiro2008beating,ecker2019overcoming}. To explore potential quantum systems suitable for realizing qudits, various photonic degrees of freedom, including polarization \cite{bogdanov2004qutrit}, orbital angular momentum \cite{karan2023postselection,cao2020distribution}, and spatial modes \cite{hiekkamaki2021high,edgar2012imaging,devaux2020imaging}, have been extensively studied for HD quantum communication, computing, and sensing. However, in practical applications such as fiber-optic communication systems, certain degrees of freedom may be vulnerable to channel noise, limiting their performance and feasibility. 

In contrast, quantum information encoded in time-frequency entanglement (TFE) is inherently more resilient to channel noise \cite{steinlechner2017distribution}. Furthermore, its high-dimensional nature makes TFE a promising candidate for overcoming practical implementation challenges. Due to these advantages, TFE has garnered significant attention in recent years \cite{brecht2015photon,gianani2020measuring,chang2025recent,yu2024time} and has been actively explored in various quantum technology applications, including quantum computing \cite{lu2023frequency,lukens2016frequency,yamazaki2023linear,cui2020high,weng2024implementation}, quantum communication \cite{zhang2014unconditional,nunn2013large,zhong2015photon,yu2025quantum,chang2023large,graffitti2020direct,niizeki2020two,khodadad2025frequency}, quantum error correction \cite{descamps2024gottesman}, and entanglement swapping \cite{merkouche2022heralding,merkouche2022spectrally,vitullo2018entanglement}. These studies further highlight the potential of TFE in HD QIP, paving the way for practical advancements in quantum technology.

Extending TFE to HD QIP requires precise control over the spectral information of photon pairs, discretization of spectral modes, and the generation of frequency-entangled qudits (FEQs). Various approaches have been proposed and experimentally demonstrated to achieve this. The most straightforward method for creating FEQs is spectral shaping via filtering techniques. For instance, external cavity filters can be used to shape the spectrum of photon pairs generated through spontaneous parametric down-conversion (SPDC), enabling the production of high-dimensional FEQs \cite{chang2023towards,chang2021648,chang2024time,cheng2023high,ikuta2019frequency,fabre2020generation,yamazaki2022massive,xie2015harnessing,niizeki2020two,chang2025recent}. Additionally, chip-based cavity spectral shaping \cite{lu2022bayesian,ortiz2021submegahertz,myilswamy2023time} and spatial-spectral mapping schemes \cite{yang2023spatial} have been explored. However, these techniques often filter out a significant portion of the photon pairs, leading to substantial losses and reduced source brightness. To mitigate these losses, FEQs can be directly generated by engineering the phase-matching conditions (PMCs) of SPDC sources, thereby shaping the joint spectral amplitude (JSA) of photon pairs \cite{shukhin2024two,morrison2022frequency}.

Another ingenious approach utilizes quantum interference to modulate the JSA of photon pairs and further manipulate its time-frequency quantum information \cite{jin2024quantum}. For instance, Hong-Ou-Mandel (HOM) interference induces interference fringes along the difference-frequency direction. This results in the JSA of anti-correlated photon pairs in a comb-like spectral structure, forming an FEQ \cite{lingaraju2019quantum,jin2016simple,chen2021temporal}. Furthermore, compared to PMC engineering, the tunability of quantum interference provides a flexible, stable, and efficient method to control various properties of time-frequency quantum information in photon pairs \cite{li2023spectrally,jin2016simple,jin2024spectrally,jin2021spectrally,chen2021temporal,hong2023fast}. 

However, the physical implications of such HOM-based FEQs may still not be fully explored, and further investigation could deepen our understanding and expand their potential applications. It is important to note that the time-frequency quantum information of photon pairs inherently belongs to a continuous-variable (CV) system \cite{fabre2022time}. While quantum interference effectively discretizes the spectral information, each discrete frequency bin may still preserve CV correlations or entanglement \textit{in the frequency domain}. When entanglement is present within individual discrete frequency bins, the generated quantum state is no longer a pure FEQ but rather a superposition of multiple TFE states. Interestingly, in the framework of CV quantum information, such a state can be interpreted as a superposition of multiple two-mode squeezed states (TMSSs) distributed at different positions in phase space \cite{fabre2022time}. This hybrid entanglement structure integrates both discrete-variable (DV) and CV characteristics within the frequency domain, opening new avenues for HD time-frequency QIP (see Sec.\ref{secV} for further discussion).

In this work, we define a novel quantum state, the \textit{hybrid frequency-entangled qudit} (HFEQ), and conduct a comprehensive theoretical and experimental study. The core of our research lies in generating and controlling its hybrid entanglement structure. More importantly, we explore its intrinsic CV entanglement, a property that, to the best of our knowledge, has not been thoroughly investigated. To generate and characterize HFEQs, we establish an integrated quantum interferometer that integrates constructive-HOM and Franson interference. HOM interference modulates and discretizes the JSA of SPDC photon pairs along the difference-frequency direction, thereby forming HFEQs. By tuning the relative time delay between the two photons, we demonstrate control over discrete dimensions, achieving $D=5,7,9,11$. To verify and quantify the CV frequency entanglement at both global and individual modes, we perform Franson two-photon interference (TPI). The TPI measurements exhibit high contrast, reaching approximately 98\% for the global state and 95\% for individual modes. Incorporating our theoretical framework, these results confirm that HFEQs possess additional entanglement beyond typical FEQs, exhibiting hybrid DV-CV time-frequency quantum entanglement. Our work introduces an innovative approach to HD time-frequency quantum information processing. Additionally, it uncovers the unexplored physics of HOM-based FEQs, offering deeper insights into the role of quantum interference in time-frequency QIP.

The paper is structured as follows: Sec. \ref{secII} presents the theoretical model, describing a setup that enables two types of interference to manipulate the SPDC JSA, thereby forming HFEQs and verifying their DV-CV characteristics. Sec. \ref{secIII} details the experimental setup and results. Sec. \ref{secIV} provides a summary of the study, while Sec. \ref{secV} concludes the findings and explores potential applications of the proposed HFEQ. Additionally, the Appendix offers further details on the experimental setup and theoretical discussion.


\begin{figure}[t]
\centering
\includegraphics[width=0.46\textwidth]{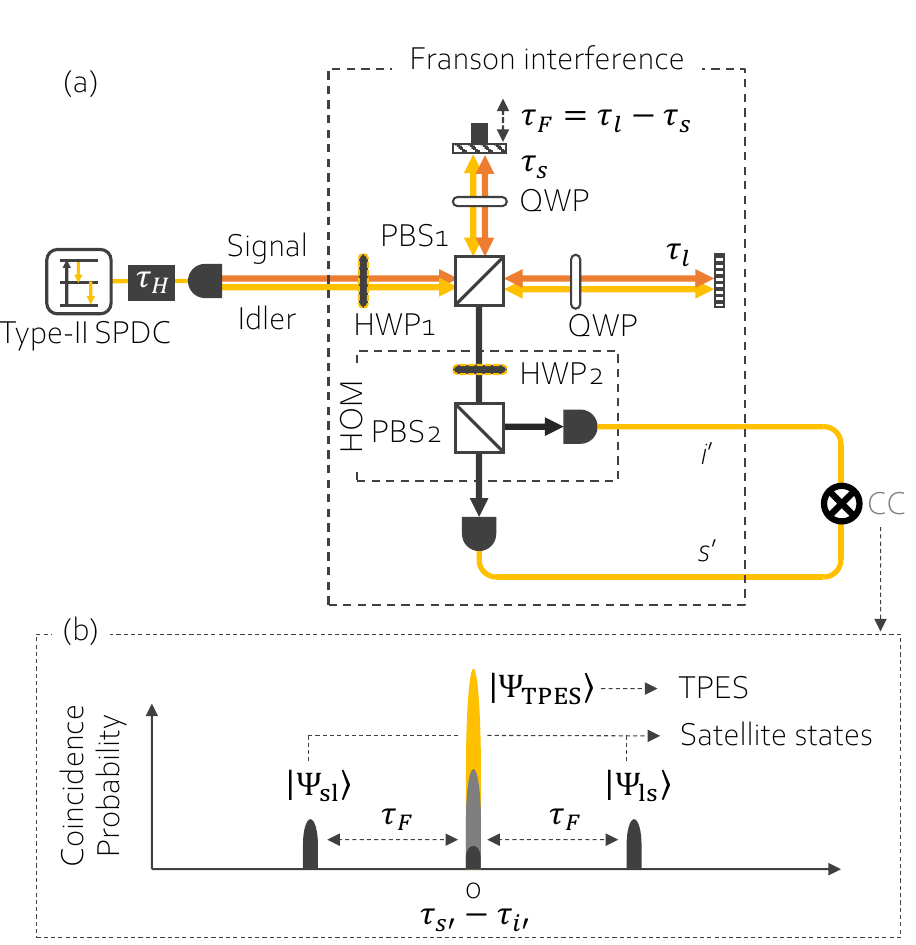}
\caption{(a) Schematic diagram of the integrated quantum interferometer. PBS: Polarizing beam splitter; HWP: Half-wave plate; QWP: Quarter-wave plate; CC: Coincidence counting measurement. (b) Schematic diagram of the coincidence counting measurement for the conditional state. $\tau_{s'}$ and $\tau_{i'}$ denotes the detected timing of $s'$-mode and $i'$-mode photons, respectively. }
\label{setup_th}
\end{figure}

\section{Theoretical model} \label{secII}


\subsection{Quantum state evolution}

The concept of the integrated quantum interferometer setup is illustrated in Fig. \ref{setup_th}. Here we start with the time-frequency entangled photon pairs that are generated by a continuous-wave (CW) pumped type-II SPDC source with a pure quantum state of
\begin{equation}
    |\Psi_{\text{SPDC}}\rangle=\int\int d\omega_sd\omega_i f(\omega_s,\omega_i)\hat{a}_H^{\dagger}(\omega_s)\hat{a}_V^{\dagger}(\omega_i)e^{i\omega_s\tau_H}\left|0\right\rangle,
\label{SPDC_QS}
\end{equation}
where $\hat{a}_{H,V}^{\dagger}(\omega)$ represents the creation operator for photons with horizontal (H) and vertical (V) polarizations at frequency $\omega$, and $\omega_{s,i}$ denotes the frequencies of the signal and idler photons, respectively. The JSA of the signal and idler photons, $f(\omega_s,\omega_i)$, characterizes the frequency entanglement of photon pairs generated by SPDC. Additionally, the joint spectral intensity (JSI), defined as $J_{I}^{\text{SPDC}}=|f(\omega_s,\omega_i)|^2$, which removes the phase information of the JSA, is a more commonly used probability distribution in the joint spectrum for analyzing the frequency correlation of photon pairs. In the following discussion, the JSA of SPDC is modeled using a common form, which is constructed from the phase-matching function of the crystal and the pump spectrum \cite{jin2013widely}, $f(\omega_s, \omega_i)=1/\sqrt{\mathcal{N}} e^{-2\ln(2)\omega_{-}^2/\Delta\omega_S^2} e^{-2ln(2)\omega_{+}^2/\Delta\omega_p^2}$, where $\omega_{\pm}=\delta\omega_s \pm \delta\omega_i$,  $\delta\omega_{s,i} = \omega_{s,i} - \omega_{s,i,0}$, $\omega_{s,i,0}$ is the central frequency of the signal and idler photons, $\Delta\omega_p$ represents the pump laser bandwidth, $\Delta\omega_S$ represents the SPDC single-photon bandwidth, and $\mathcal{N}$ is the normalized constant.  

In Eq. (\ref{SPDC_QS}), a time delay of $\tau_H$ is applied after $\hat{a}_H^{\dagger}(\omega_s)$ to introduce the phase factor $e^{i\omega_s\tau_H}$.  Generally, this delay represents the intrinsic time difference between the signal and idler photons, arising from the different refractive indices associated with the two orthogonal polarizations. To study the influence of this parameter, we assume that $\tau_H$ can be controlled by an additional mechanism. This parameter is crucial in experiments aimed at manipulating frequency entanglement and generating HFEQs, and its effects will be explored further.

The generated photon pairs are directed into a polarization-based Franson interferometer to measure TPI, which is used to verify the frequency correlation of the photon pairs. Unlike a conventional Franson interferometer, our setup employs a single unbalanced Michelson interferometer to demonstrate Franson’s TPI \cite{park2018time}. In this configuration, the signal and idler photons are initially polarized to linear diagonal states ($\pm45^\circ$) using HWP1. PBS1 then evenly distributes the photon pairs between the interferometer’s long and short arms. The length difference between these arms exceeds the single-photon coherence length of SPDC, which prevents single-photon interference while preserving TPI. In each arm, a QWP is placed along the optical path, together with the mirror at the end, this forms a double-pass configuration. This design allows photons passing through the long and short arms to be recombined by PBS1 and output through the same port. After considering the setup above, the field operators can be replaced in this phase as $\hat{a}_{H/V}^{\dagger}(\omega)\rightarrow \left[\hat{a}_V^{\dagger}(\omega)e^{i\omega\tau_l}\pm\hat{a}_H^{\dagger}(\omega)e^{i\omega\tau_s} \right]/\sqrt{2}$, where $\tau_s$ and $\tau_l$ represent the travel times through the short and long arms, respectively. 

To observe TPI through coincidence counting, a splitting mechanism is necessary to direct the entangled photon pairs into different modes. To achieve this, an HWP2 is set at an angle of $22.5^\circ$, rotating the H- and V-polarized photons to $\pm45^\circ$. PBS2 then splits the biphotons into two distinguish outputs (denoted as $s'$- and $i'$-mode, see Fig. \ref{setup_th}). As a result, the field operators evolve according to the relation $\hat{a}_{H/V}^{\dagger}(\omega)\rightarrow [\hat{a}_{s'}^{\dagger}(\omega)\pm\hat{a}_{i'}^{\dagger}(\omega)]/\sqrt{2}$. With this relation and also considering the coincidence counting measurement, the final conditional quantum state (without renormalization and removing the global phase) in $s'$- and $i'$-mode is given by
\begin{widetext}
\begin{equation}
\begin{aligned}
|\Psi_{\text{out}}\rangle=\frac{1}{4}\int\int d\omega_sd\omega_i f(\omega_s,\omega_i)e^{i\omega_s\tau_H}\times
&\left\{\begin{array}{l}
    \left[ \hat{a}_{s'}^{\dagger}(\omega_s)\hat{a}_{i'}^{\dagger}(\omega_i)+\hat{a}_{s'}^{\dagger}(\omega_i)\hat{a}_{i'}^{\dagger}(\omega_s) \right]\times\left[e^{i(\omega_s+\omega_i)\tau_l}+e^{i(\omega_s+\omega_i)\tau_s} \right]+ \\
    \left[ \hat{a}_{s'}^{\dagger}(\omega_s)\hat{a}_{i'}^{\dagger}(\omega_i)e^{i(\omega_s\tau_l+\omega_i\tau_s)}-\hat{a}_{s'}^{\dagger}(\omega_i)\hat{a}_{i'}^{\dagger}(\omega_s)e^{i(\omega_s\tau_s+\omega_i\tau_l)} \right]+\\
   \left[ \hat{a}_{s'}^{\dagger}(\omega_s)\hat{a}_{i'}^{\dagger}(\omega_i)e^{i(\omega_s\tau_s+\omega_i\tau_l)}-\hat{a}_{s'}^{\dagger}(\omega_i)\hat{a}_{i'}^{\dagger}(\omega_s)e^{i(\omega_s\tau_l+\omega_i\tau_s)} \right]
\end{array}\right\}\left|0\right\rangle,
\label{phi_4}
\end{aligned}
\end{equation}
\end{widetext}
where $\hat{a}_{s',i'}^{\dagger}(\omega)$ is the creation operator for the $s'$ and $i'$-mode, respectively. 

In Eq. (\ref{phi_4}), the physical meaning of the quantum state can be reinterpreted based on the travel time of each term. For the first term (denoted as $|\Psi_{\text{TPES}}\rangle$, referred to as the \textit{two-photon entangled state}, TPES), photons with frequencies $\omega_s$ and $\omega_i$ simultaneously pass through either the long or short arms of the interferometer. As a result, they become indistinguishable in coincidence count measurements, forming an entangled state with TPI effects \cite{franson1989bell}.

The second term describes a scenario in which a photon traveling through the long arm is always detected in the  $s'$-mode (photons with a delay of $\tau_l$ are detected in the $s'$-mode), while photons with a delay of $\tau_s$ are detected in the $i'$-mode (denoted as $|\Psi_{\text{ls}}\rangle$). The third term (denoted as $ |\Psi_{\text{sl}}\rangle$) corresponds to the inverse scenario of the second term. Both states are referred to as \textit{satellite states}, which are without TPI effect. Based on the travel times of these states, they can be identified in the coincidence count measurement, as illustrated in Fig. \ref{setup_th}(b).


To comprehensively investigate the generation and verification of HFEQs using the integrated interferometer, we focus on TPES, which allows us to demonstrate that the interferometer provides complete control over frequency entanglement through quantum interference (The impact of the interferometer on the satellite states will be discussed in Appendix. \ref{satellite_states}).

Now, let us consider the TPES at the output of the interferometer. This quantum state can be redefined in terms of the frequency basis observed in the $s'$- and $i'$-modes, as
\begin{equation}
\begin{aligned}
|\Psi_{\text{TPES}}\rangle=&\int\int d\Omega_sd\Omega_iJ^{\text{TPES}}_A(\Omega_s,\Omega_i)|\Omega_s,\Omega_i\rangle,
\label{QS_llss}
\end{aligned}
\end{equation}
where $\Omega_{s,i}$ denotes the frequencies in the $s'$ and $i'$ modes, $|\Omega_s,\Omega_i\rangle=\hat{a}^{\dagger}_{s'}(\Omega_s)\hat{a}^{\dagger}_{i'}(\Omega_i)|0\rangle$, and $J^{\text{TPES}}_A(\Omega_s,\Omega_i)$ is the JSA of TPES that given by 
\begin{equation}
\begin{aligned}
J^{\text{TPES}}_A(\Omega_s,\Omega_i)&=\langle\Omega_s,\Omega_i|\Psi_{\text{TPES}}\rangle\\
&=\frac{1}{2}\left[f(\Omega_s,\Omega_i)+f(\Omega_i,\Omega_s)e^{-i\Omega_{-}\tau_H}\right]\\
&\times\cos\left(\Omega_{+}\frac{\tau_F}{2}\right)e^{i\Omega_{s}(\tau_l+\tau_s+\tau_H)}.
\label{JSA_llss}
\end{aligned}
\end{equation}
where $\Omega_{\pm}=\Omega_{s}\pm\Omega_{i}$ denotes the sum and difference frequency of coordinates in the joint spectrum. $\tau_F=\tau_l-\tau_s$ is the time difference between the long and short arms. 

Eq. (\ref{JSA_llss}) reveals several intriguing phenomena. First, we observe the presence of two SPDC JSAs, where they exchange frequency components and exhibit a phase difference due to the previously introduced time delay and the frequency difference, $\Omega_{-}\tau_H$. This frequency exchange between the two modes corresponds to HOM interference. Furthermore, the positive sign indicates that the interference is constructive rather than the conventional destructive HOM interference \cite{jin2016simple,kim2017two}.

This phenomenon occurs because, in the state $|\Psi_{\text{TPES}}\rangle$, both photons travel through either the long or short arm simultaneously. As a result, they retain the same polarization and re-encounter each other at PBS2, where they interfere constructively. This behavior contrasts with that of the satellite states, which undergo destructive HOM interference at PBS1 (see Appendix \ref{satellite_states} for details).

Another crucial phenomenon described in Eq. (\ref{JSA_llss}) is its capability to reveal the effects of Franson's TPI. Interestingly, this effect is global, as it does not depend on whether the original SPDC JSA is degenerate. Moreover, since the interference occurs along the sum frequency direction, $\Omega_{+}$, this characteristic provides a direct way to analyze the frequency anti-correlation of the photon pairs. A more detailed discussion on this topic will be presented in the following section.

Naturally, the occurrence of the aforementioned HOM effect depends on the extent to which the SPDC JSA retains overlap in the joint spectrum after the frequency exchange process in Eq.\ref{JSA_llss}. To highlight the impact of the HOM effect on the quantum states, we consider the degenerate and symmetric conditions for SPDC operation, i.e., $f(\Omega_s, \Omega_i) = f(\Omega_i, \Omega_s)$. (Appendices \ref{satellite_states} and \ref{degenerate} provide additional studies on the relationship between SPDC degeneracy and quantum interference.) Under these conditions, Eq. (\ref{JSA_llss}) can be further simplified and expressed as:
\begin{equation}
\begin{aligned}
J^{\text{TPES}}_A(\Omega_s,\Omega_i) =f(\Omega_s,\Omega_i)\underbrace{\cos\left(\Omega_{+}\frac{\tau_F}{2}\right)}_{\text{Franson}}\underbrace{\cos\left(\Omega_{-}\frac{\tau_H}{2}\right)}_{\text{HOM}}.
\label{JSA_llss2}
\end{aligned}
\end{equation}
In Eq. (\ref{JSA_llss2}), we can more clearly observe how Franson and HOM interference, respectively, influence the JSA of SPDC. By independently controlling two different time delays of $\tau_F$ and $\tau_H$, the SPDC JSA can be effectively modulated along both the difference-frequency direction and the sum-frequency direction simultaneously, which provides a unique mechanism to manipulate the quantum information of $|\Psi_{\text{TPES}}\rangle$ in the frequency domain.

In the following discussion, we will separately describe the effects of two types of interference on the modulation of the frequency quantum information. Ultimately, we will demonstrate the advantages of this integrated interferometer in manipulating and verifying the frequency information of the photon pairs.

\begin{figure}[t]
\centering
\includegraphics[width=0.485\textwidth]{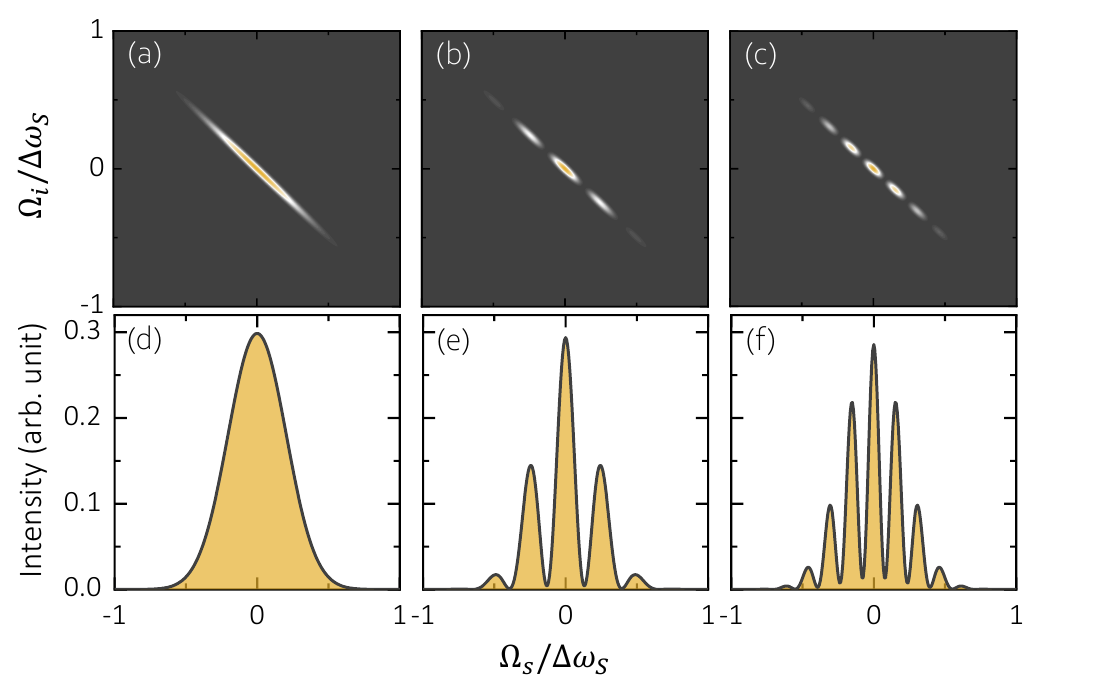}
\caption{Demonstration of a series of highly anti-correlated TPES JSIs under constructive HOM interference. In (a)-(c), we examine a degenerate JSI with a series of HOM delay times $\tau_H$ of $0$, $12/\Delta\omega_S$, and $20/\Delta\omega_S$, respectively. (d)-(f) show the reduced signal mode spectrum of (a)-(c). The pump laser bandwidth is set to $\Delta\omega_p=\Delta\omega_S/20$ for all cases.
}
\label{HOM_comb}
\end{figure}

\subsection{Generation of HFEQs: HOM interference}\label{thoeory_HOM}

At the beginning of this discussion, we first focus on the effect of HOM interference on TPES. Here, we temporarily ignore the effect of Franson interference in Eq. (\ref{JSA_llss2}) and consider a common condition where the input photon pairs are highly frequency anti-correlated ($\Delta\omega_p \ll \Delta\omega_S$). The JSI of TPES, $J_I^{\text{TPES}}(\Omega_s,\Omega_i) = |J_A^{\text{TPES}}(\Omega_s,\Omega_i)|^2$, after interference under different HOM delay conditions is presented in Figs. \ref{HOM_comb}(a)–\ref{HOM_comb}(c). With the introduction of HOM interference, we observe that when $\tau_H = 0$, the fully constructive HOM interference does not alter the original SPDC JSI and preserves its CV correlation. On the other hand, when $\tau_H \neq 0$, the JSI of TPES is modulated into a comb-like structure, indicating that TPES has the potential to evolve into a high-dimensional frequency-discretized entangled state. In the language of DV quantum information, such a state can be referred to as the FEQs \cite{hong2023fast,jin2016simple,chen2021temporal}. Furthermore, as $\tau_H$ increases, Figs. \ref{HOM_comb}(a)-\ref{HOM_comb}(c) demonstrate that the setup enables the manipulation of the dimensionality of FEQs. For instance, in Fig. \ref{HOM_comb}(b), the estimated dimension is $D=5$, while in Fig. \ref{HOM_comb}(c), it is approximately $D=7$. 

However, from our perspective, the FEQs generated via HOM interference not only offer advantages for DV quantum information applications but also hold significant potential for CV QPI. A careful observation of the distribution within individual frequency bins in Fig. \ref{HOM_comb}(b) and \ref{HOM_comb}(c) reveals a certain degree of frequency correlation in each bin, implying it remains more interesting quantum information properties to be explored in FEQs generated by HOM interference.

Therefore, unlike conventional FEQs, which are typically regarded as a discrete superposition of correlated frequency modes \cite{jin2016simple,chen2021temporal},  
\begin{equation}
\begin{aligned}
|\Psi_{\text{FEQ}}\rangle = \sum_{n}^{D}a_n|\omega_{n},\omega_{D-n}\rangle,
\label{FEQs_state}
\end{aligned}
\end{equation}
where $a_n$ is the amplitude, $\omega_{n}$ is the frequency of the $n$th bin, and $D$ is the dimensionality of FEQs. In contrast, the generated quantum states by HOM interference should be more accurately described as  
\begin{equation}
\begin{aligned}
&|\Psi_{\text{HFEQ}}\rangle \\
&\approx \sum_{n=-\frac{(D-1)}{2}}^{\frac{(D-1)}{2}}a_n\int\int d\Omega_sd\Omega_iJ_{A,n}(\Omega_{s,n},\Omega_{i,n})|\Omega_{s},\Omega_{i}\rangle,
\label{HFEQs_state}
\end{aligned}
\end{equation}
where $D$ is the dimensionality of the generated states, which is always an odd number due to constructive HOM interference. $\Omega_{s(i),n}=\Omega_{s(i)}-\Omega_{0,n}$, and $\Omega_{0,n}$ represents the central frequency of each frequency bin of the generated states. $J_{A,n}(\Omega_{s,n},\Omega_{i,n})$ denotes the JSA for the $n$th frequency bin with CV entanglement, which cannot be factorized into two independent functions, i.e.,  $J_{A,n}(\Omega_{s,n},\Omega_{i,n})\neq f_s(\Omega_{s,n}) \times f_i(\Omega_{i,n})$. 

From Eqs.\ref{FEQs_state} and \ref{HFEQs_state}, we can observe that both $|\Psi_{\text{FEQ}}\rangle$ and $|\Psi_{\text{HFEQ}}\rangle$ exhibit discrete entanglement structures with a dimensionality $D$. However, compared to $|\Psi_{\text{FEQ}}\rangle$, the state $|\Psi_{\text{HFEQ}}\rangle$ not only possesses a DV structure but also exhibits intrinsic CV entanglement within each individual frequency bin. This results in a superposition of multiple time-frequency CV-entangled states, leading to a higher degree of entanglement than $|\Psi_{\text{FEQ}}\rangle$. Due to its unique hybrid entanglement structure that integrates both DV and CV features, we refer to $|\Psi_{\text{HFEQ}}\rangle$ as \textit{hybrid frequency-entangled qudits} (HFEQs). 

This distinct time-frequency entanglement structure not only provides deeper insights into the fundamental nature of the HOM-based FEQs but also offers a novel resource for time-frequency QIP, which will be further discussed in Sec.\ref{secV}.

\subsection{Characterization of HFEQs}

To emphasize the advantage of the entanglement of HFEQs over FEQs, we formulate the indicators to characterize HFEQs in this section. For the DV property, the dimensionality $D$ represents the size of the Hilbert space of HFEQs. For the CV property, one can focus on the individual frequency bin to verify its frequency entanglement through Franson interference. 

Moreover, to further demonstrate the higher entanglement of HFEQs than FEQs, we can employ the Schmidt decomposition method to further expand the JSA of both quantum states using two orthogonal bases and estimate their Schmidt numbers, providing a quantitative measure of entanglement \cite{li2023pure}. For an FEQ with dimensionality $D$, the maximum Schmidt number is $D$ when each frequency bin is uniformly distributed. However, for an HFEQ with the same dimensionality, each independent frequency bin possesses additional entanglement. As a result, these contributions enhance the global frequency entanglement of the HFEQ, leading to a Schmidt number of HFEQ, $K_F$, possibly greater than $D$ for an HFEQ of dimensionality $D$.




To measure the aforementioned characteristics, we propose a simple and experimentally feasible scheme for estimating these key parameters. First, the dimensionality of the HFEQs can be determined by analyzing the spectral structure of single photons under HOM interference. Simultaneously, the HOM interference visibility of the single-photon spectrum provides an estimate of the Schmidt number of the HFEQs. Combined with the implementation of Franson-type TPI, this approach further verifies the frequency correlations within the HFEQs, enabling a more comprehensive evaluation of their key performance parameters. In the following discussion, we present the details of this method.


\subsubsection{Verification of dimensionality and Schmidt number}

To estimate the dimensionality of HFEQs, a straightforward approach is to use one photon of the photon pair as a heralding signal and perform a conditional spectral measurement on the other photon. By analyzing the profile of this single-mode spectrum, the dimensionality of HFEQs can be determined. Theoretically, we can use Eq. (\ref{JSA_llss2}) to simulate the spectral measurement results corresponding to this process. If the idler photon is used as the heralding signal in the measurement, the single-mode spectrum of the signal photon, $S_s(\Omega_s)$, can be expressed as
\begin{equation}
\begin{aligned}
S_s(\Omega_s,\tau_H)=\int d\Omega_iJ^{\text{TPES}}_I(\Omega_s,\Omega_i).
\label{SPS}
\end{aligned}
\end{equation}
With Eq. (\ref{SPS}), the single-photon spectra corresponding to Figs. \ref{HOM_comb}(a)-\ref{HOM_comb}(c) are presented in Figs. \ref{HOM_comb}(d)-\ref{HOM_comb}(f). These results reveal the single-mode spectra of different TPES and allow us to identify the dimensionality of their corresponding HFEQs. 

For the estimation of entanglement in HFEQs, interestingly, although the above approach only provides \textit{partial} information about the JSI of HFEQs, some of these JSI correlations can still be inferred from certain effects observed in Fig. \ref{HOM_comb}(e) and Fig. \ref{HOM_comb}(f). To demonstrate this, let us now focus on Fig. \ref{HOM_comb}(f). We observe that the comb-like spectral structure induced by HOM interference does not exhibit perfect contrast. This imperfection arises from the finite pump bandwidth, which reduces the two-photon frequency correlation in the SPDC JSA and consequently affects the interference contrast in the single-mode spectrum. Therefore, the interference contrast actually reflects the frequency correlation of HFEQs to a certain extent. This effect suggests an elegant approach for estimating entanglement in HFEQs: by analyzing the HOM interference visibility in the single-mode spectrum, one can infer the global frequency entanglement within the HFEQs and further estimate its Schmidt number $K_F$.

\begin{figure}[t]
\centering
\includegraphics[width=0.485\textwidth]{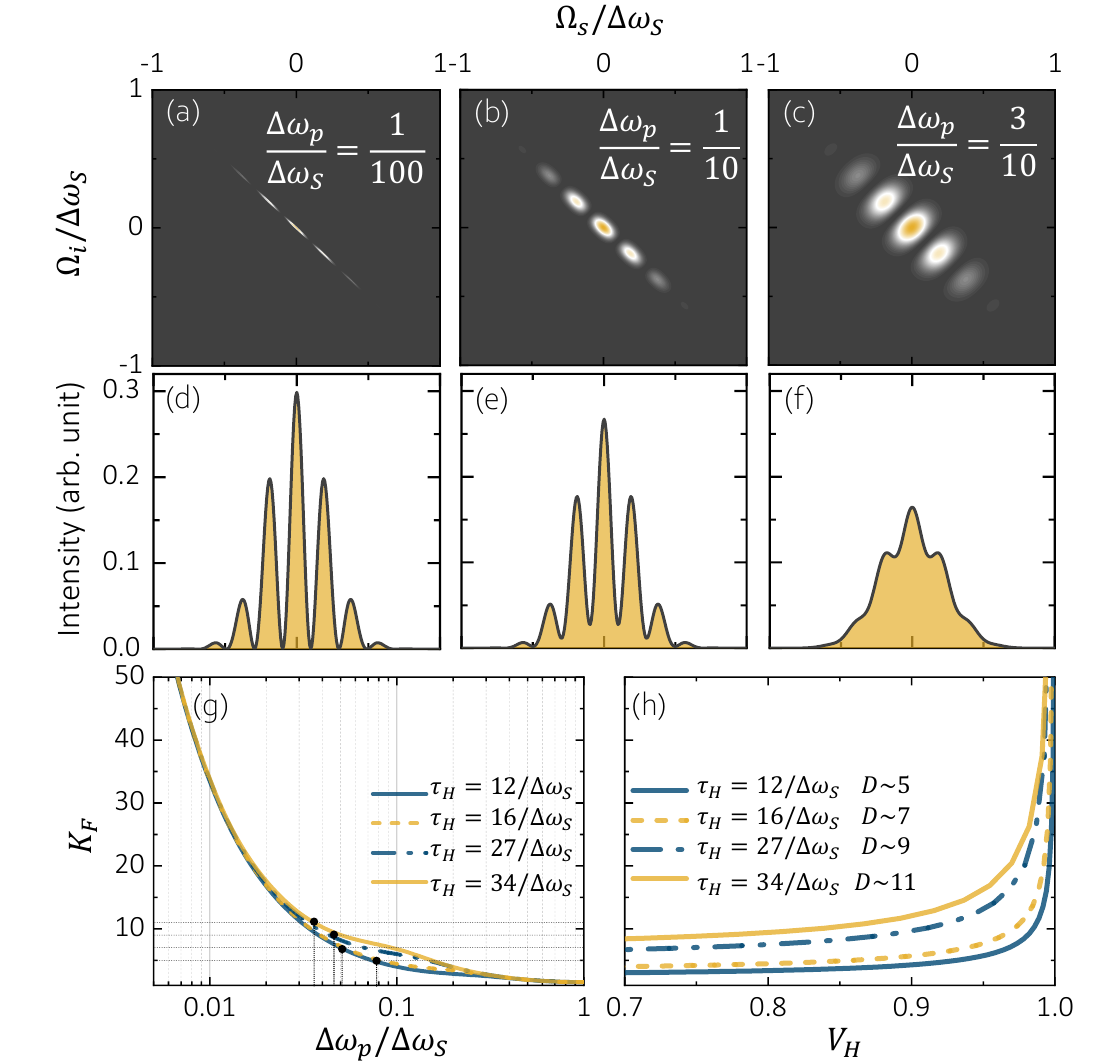}
\caption{Comparison of HOM visibility and frequency correlation of HFEQs. (a)-(c) The JSIs of HFEQs with $\tau_H=16/\Delta\omega_S$ under a series of pump bandwidths (as shown in labels). (d)-(f) The corresponding single-photon spectrum of (a)-(c), respectively. (g) The relationship between the Schmidt number of HFEQs, $K_F$, and pump bandwidth. (h) The relationship between $K_F$ and HOM visibility, $V_H$.}
\label{HOM_visibility}
\end{figure}

To investigate the relationship between $K_F$ and the visibility of the HOM interference in the single-mode spectrum, we vary the pump bandwidth, $\Delta\omega_p$, to control the frequency correlation of SPDC. The resulting HOM interference patterns, shown in Figs. \ref{HOM_visibility}(a)-\ref{HOM_visibility}(f) exhibit a gradual decrease in visibility as the pump bandwidth increases due to the weakened frequency correlation in the SPDC JSI.

Furthermore, Fig. \ref{HOM_visibility}(g) illustrates the relationship between the pump bandwidth and $K_F$ for a series of values of $\tau_H$. As expected, a narrower pump bandwidth results in a higher $K_F$. In addition, as $\tau_H$ increases, the generated HFEQs exhibit higher dimensions, leading to an increase in $K_F$ at the same $\Delta\omega_p/\Delta\omega_S$. Notably, the parameter settings for the bandwidth of $\tau_H\Delta\omega_S=12,16,27,$ and $34$ in Fig. \ref{HOM_visibility}(g) approximately correspond to $D=5,7,9,$ and $11$, respectively, which impose upper limits on the FEQs maximum Schmidt number of $D$. As $\Delta\omega_p/\Delta\omega_S$ decreases, $K_F$ can easily surpass the limitation of FEQ's maximum Schmidt number, further demonstrating the advantage of HFEQs in terms of entanglement degree. 

Next, to evaluate the HOM visibility, we first extract the interference pattern from the single-mode spectrum. This is accomplished by normalizing the single-mode spectrum with respect to the original single-mode spectrum of SPDC.
\begin{equation}
\begin{aligned}
S_{H}(\Omega_s,\tau_H)=\frac{S_{s}(\Omega_s,\tau_H)}{\int d\Omega_i|f(\Omega_s,\Omega_i)|^2}.
\end{aligned}
\end{equation}
Then, the HOM visibility could be calculated as 
\begin{equation}
\begin{aligned}
V_{H}=\frac{S_{H}^{\max}(\Omega_s,\tau_H)-S_{H}^{\min}(\Omega_s,\tau_H)}{S_{H}^{\max}(\Omega_s,\tau_H)+S_{H}^{\min}(\Omega_s,\tau_H)},
\label{V_H}
\end{aligned}
\end{equation}
where $S_{H}^{\max}(\Omega_s,\tau_H)$ and $S_{H}^{\min}(\Omega_s,\tau_H)$ denote the maximum and minimum values of $S_{H}$, respectively. 

Fig. \ref{HOM_visibility}(h) reveals a clear relationship between the frequency correlation of HFEQs and the visibility of HOM interference. Specifically, a higher experimentally observed $V_H$ corresponds to a higher Schmidt number $K_F$. Moreover, Fig. \ref{HOM_visibility}(h) also illustrates the impact of varying $\tau_H$. For a fixed $K_F$, a smaller $\tau_H$ leads to higher HOM interference visibility. This occurs because a smaller $\tau_H$ reduces the HOM modulation frequency applied to the TPES, resulting in lower-dimensional HFEQs. Consequently, the frequency bins become more distinguishable in the single-mode spectrum, and vice versa. In practice, a larger $\tau_H$ imposes a more stringent requirement on interference contrast, thereby enhancing the accuracy in determining $K_F$. This behavior provides a novel approach to infer the strength of frequency entanglement in generated HFEQs by monitoring the HOM interference visibility of TPES.

Of course, certain limitations still exist in this method. First, in practical experiments, the measurement of the single-mode spectrum may have a finite resolution, which imposes an upper bound on the estimated HOM visibility (Appendix.\ref{jitter_app}). Therefore, the inferred $K_F$ in the real experiment represents a lower bound on the actual entanglement strength of the HFEQs. Second, measuring the spectrum of only one mode of HFEQs provides only partial information about the quantum state, which imposes certain limitations on this approach. For instance, consider an initial photon pair without frequency entanglement, where one photon undergoes spectral modulation (e.g., through a cavity filter) while the other photon retains its original spectrum. In this case, we may still observe a clear contrast in the single-mode spectrum, even though the quantum state is not truly an HFEQ. Therefore, to determine whether the prepared HFEQs exhibit genuine frequency correlation, complementary approaches must be explored to obtain more comprehensive information about the quantum state.

Fortunately, the integrated interferometer presented in this work offers another essential quantum interference effect: Franson's TPI, which provides interference along the sum-frequency direction. This additional interference mechanism grants us an extra degree of freedom to explore the frequency correlation of the prepared HFEQs. Moreover, it introduces an intriguing approach for manipulating the time-frequency quantum information of SPDC photon pairs, which will be discussed in the next section.

\begin{figure}[t]
\centering
\includegraphics[width=0.485\textwidth]{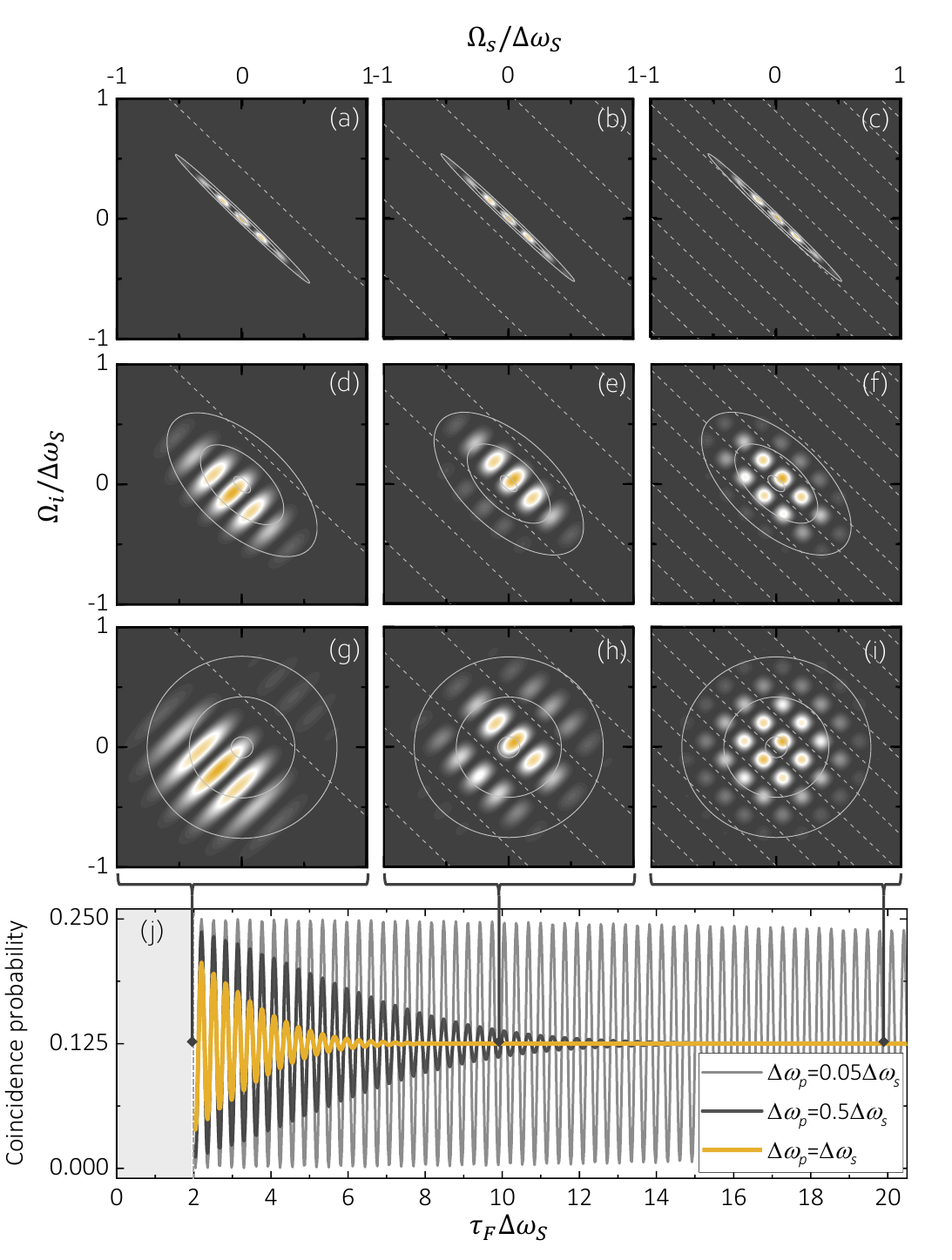}
\caption{Simulations of different bandwidth-pumped SPDC JSIs under HOM and Franson interference for various Franson delays $\tau_F$. In all cases, the HOM delay is set to $\tau_H=20/\Delta\omega_S$. Each row of figures corresponds to the same pump bandwidth: (a)-(c) $\Delta\omega_p=0.05\Delta\omega_S$, (d)-(f) $\Delta\omega_p=0.5\Delta\omega_S$, and (g)-(i) $\Delta\omega_p=\Delta\omega_S$. Each column of figures corresponds to the same Franson delay: (a)-(g) $\tau_F=2/\Delta\omega_S$, (b)-(h) $\tau_F=10/\Delta\omega_S$, and (c)-(i) $\tau_F=20/\Delta\omega_S$. The white dashed line represents the trough location of $\cos^2(\Omega_{+}\tau_F/2)$, and the white solid line denotes the distribution of origin SPDC JSI. (j) shows the TPI fringes of different bandwidth-pumped SPDCs under HOM and Franson interference. The gray area represents the signal-photon interference region which will be avoided in the Franson interferometer. The pump frequency is set at $\omega_p=20\Delta\omega_s$ to clearly demonstrate the Franson interference fringe. Notably, (f) and (i) present particular grid structures in the joint spectrum, which will be discussed in Appendix.\ref{GKP_states}. }
\label{JSI_series2}
\end{figure}

\subsubsection{Verification of frequency correlation: Franson interference}\label{thoeory_TPI}

Now, we reintroduce Franson interference into the JSA of TPES in Eq. (\ref{JSA_llss2}) to further explore the nature of the generated HFEQs. To illustrate the unique impact of Franson interference on the frequency characteristics of TPES and to investigate the relationship between this interference and the frequency correlation of HFEQs, Fig. \ref{JSI_series2} presents the $J_I^{\text{TPES}}$ under varying pump bandwidths and a series of $\tau_F$ values, with a nonzero HOM delay of $\tau_H=20/\Delta\omega_s$ to modulate the TPES into a potential HFEQ. Additionally, Fig. \ref{JSI_series2}(j) simulated coincidence probability for characterizing the TPES in the integrated interferometer, $P(\tau_F)$, as a function of $\tau_F$ for different pump linewidth conditions, providing insights into the relationship between the Franson TPI and the strength of frequency correlation. Here, $P(\tau_F)$ is calculated as
\begin{equation}
\begin{aligned}
P(\tau_F)=\int\int d\Omega_sd\Omega_i J^{\text{TPES}}_I(\Omega_s,\Omega_i).
\end{aligned}
\end{equation}

We now examine the impact of Franson interference on the JSI of HFEQs. As shown in Figs. \ref{JSI_series2}(a)-\ref{JSI_series2}(c), the strong frequency correlations result in a narrow distribution along the sum-frequency direction in the JSI. This narrow distribution limits its ability to span multiple periods of Franson sum-frequency interference. Therefore, as $\tau_F$ increases, the coincidence probability $P(\tau_F)$ of the frequency highly anti-correlated JSI becomes strongly dependent on $\tau_F$, leading to high-contrast TPI, as illustrated in Fig. \ref{JSI_series2}(j).

In Figs. \ref{JSI_series2}(d)-\ref{JSI_series2}(f), increasing the pump linewidth reduces frequency correlations. The weaker correlations allow JSI to cover more periods of Franson sum-frequency interference. Therefore, as $\tau_F$ increases, the variations of $P(\tau_F)$ diminish, leading to a pronounced decrease in TPI contrast, as shown by the black line in Fig. \ref{JSI_series2}(j).

Under conditions where the SPDC photon pairs lack frequency correlation (e.g., Figs. \ref{JSI_series2}(g)- \ref{JSI_series2}(i), $\Delta\omega_p=\Delta\omega_S$), the TPI interference contrast decreases more significantly as $\tau_F$ increases. While theoretically, reducing $\tau_F$ can maintain high TPI contrast even for weakly correlated photon pairs, the Franson interferometer requires enough path-length difference to avoid single-photon interference (gray area in Fig. \ref{JSI_series2}(j)), which limits the minimum value of $\tau_F$. Thus, for uncorrelated photon pairs, it is hard to observe a high TPI contrast. 

The above examples demonstrate that the visibility of Franson’s TPI is intrinsically linked to the frequency correlations of the HFEQs under investigation and offer a practical method to evaluate HFEQ frequency correlations through TPI visibility measurements. The high visibility of TPI represents highly anti-correlated photon pairs. Furthermore, TPI visibility also serves as an indicator of photon pair nonlocality. Unentangled photon pairs generally cannot exceed a TPI contrast of 70.7\%, as bounded by the  Clauser–Horne–Shimony–Holt (CHSH) Bell inequality \cite{tittel1999long}.

Another key characteristic of HFEQs—the frequency correlation within individual frequency bins, which also describes the CV correlation of HFEQs—can be tested using this method. This can be achieved by selectively filtering specific frequency ranges from the single-mode spectrum of HFEQs to extract the corresponding JSI, $|J_{A,n}(\Omega_{s,n},\Omega_{i,n})|^2$. The frequency correlation can then be further verified by measuring the TPI visibility.

Notably, the frequency structure of HFEQs originates from the modulation of $\Omega_{-}$ in HOM interference. Therefore, when $J_{I}^{\text{SPDC}}$ exhibits strong anti-correlation, each frequency bin retains the same two-photon bandwidth as $J_{I}^{\text{SPDC}}$. Consequently, the TPI measurements for each frequency bin yield identical interference fringes and contrast. However, it is important to note that the filtering process reduces the single-photon bandwidth of the global quantum state. As a result, the minimum allowable value of $\tau_F$ in Franson interference increases compared to the unfiltered case. Therefore, even if the two-photon bandwidth is narrow, achieving high TPI contrast remains challenging unless strong frequency correlations exist within each bin.

\subsection{Summary of theoretical model}

Before delving into the experimental details, we make a brief summary of our theoretical proposal for the generation and verification of HFEQs. Based on theoretical insights, the integrated interferometer generates a TPES from the input SPDC photon pairs. This state exhibits both Franson TPI and HOM interference effects. When the initial SPDC JSA is strongly anti-correlated, HOM interference efficiently transforms the state into an HFEQ.

To fully characterize HFEQs, we propose first measuring Franson TPI to confirm their frequency correlation. Once verified, the structure of the single-mode spectrum determines the dimensionality of HFEQs. To further analyze the frequency correlation within individual frequency bins, specific bins can be filtered from the single-mode spectrum, followed by TPI measurements to verify their internal correlations. Finally, the HOM visibility in the single-mode spectrum provides crucial information for estimating the Schmidt number of HFEQs, $K_F$, to describe the global frequency entanglement of the states. These analyses offer a comprehensive characterization of the prepared HFEQs.

On the other hand, the proposed verification method relies on correlation measurements rather than amplitude measurements, which means it does not capture the phase information. For instance, a frequency-correlated comb-like mixed state can also exhibit high TPI visibility in a Franson interferometer. However, it is crucial to emphasize that in our proposed scheme the HFEQ is generated through HOM interference. Furthermore, HOM interference induces the high-contrast comb-like spectral structure, which can only be realized in a pure state \cite{jin2016simple}. Any mixed-state component would reduce the contrast of this structure. Therefore, if the experimental results demonstrate high-contrast HOM interference, it strongly suggests that the prepared HFEQ is predominantly a pure state, as described by Eq. (\ref{HFEQs_state}).


\begin{figure*}[ht]
\centering
\includegraphics[width=0.95\textwidth]{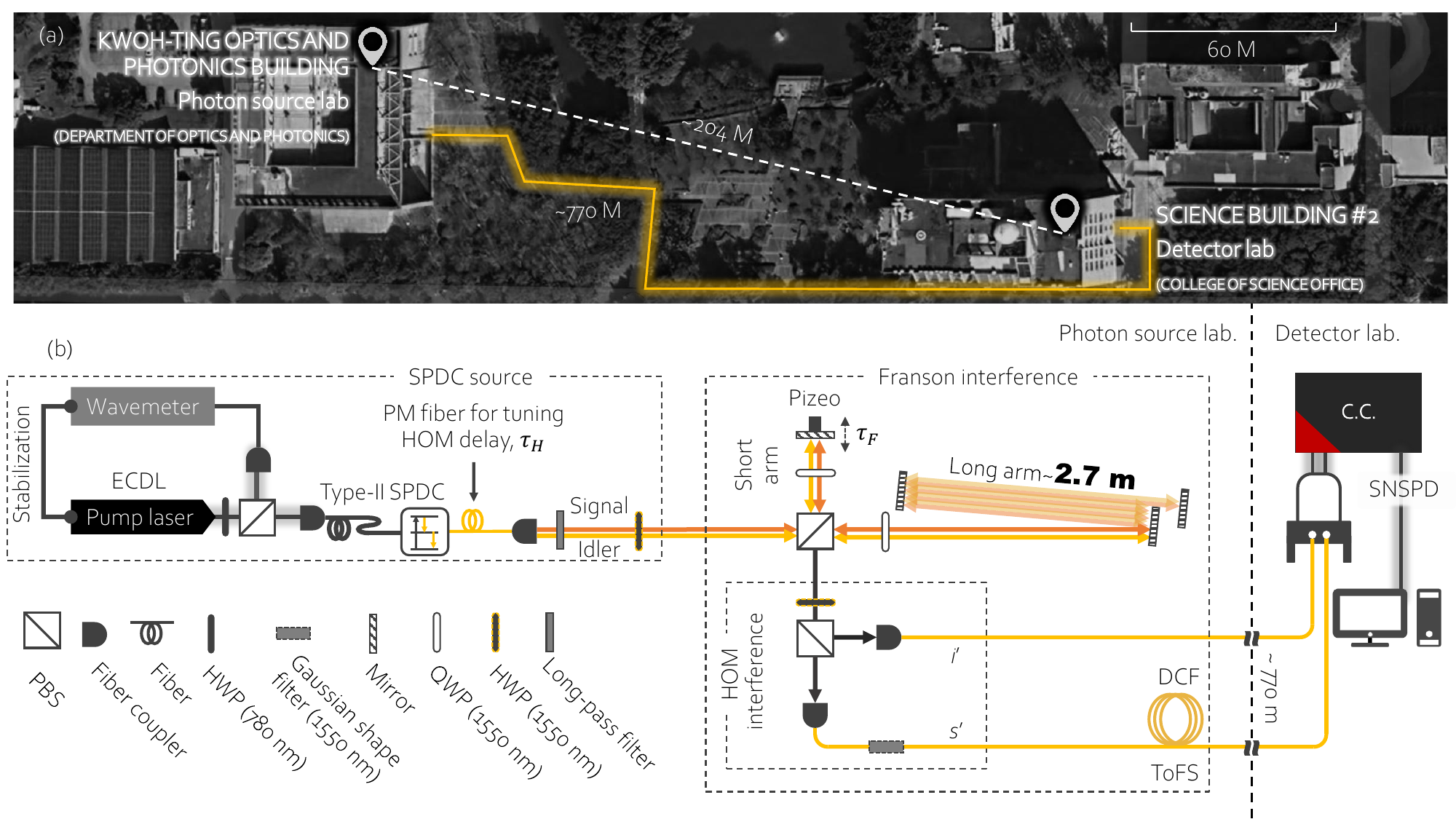}
\caption{(a) Fiber network (yellow line) in NCU campus. (b) Schematic diagram of the experimental setup. DCF: Dispersion compensating fiber. ToFS: Time-of-flight spectrometer. C.C.: coincidence counting.}
\label{setup_detail}
\end{figure*}

\section{Experiment}\label{secIII}

In this section, we move into the experimental phase, detailing the setup, results, and discussion. We aim to validate the integrated interferometer. By leveraging HOM interference, we generate HFEQs and verify their dimensionality and entanglement properties using the Franson interference. Additionally, we demonstrate the distribution of HFEQs over a practical fiber network, highlighting the robustness of time-frequency quantum information.

\subsection{Experimental setup}

To harness and distribute HFEQs using the integrated interferometer, two separate laboratories,  Photon Source Lab (PSL) and Detector Lab (DL), are set up and connected via pre-existing campus optical fibers spanning 770 meters, with the buildings separated by approximately 204 meters, as shown in Fig. \ref{setup_detail}(a). The PSL houses the integrated interferometer and photon source, while a superconducting nanowire single-photon detector (SNSPD, Single Quantum) is installed in the DL's server room. This configuration provides a practical setup for long-distance HFEQ transmission and serves as a foundation for future quantum communication experiments.

In Fig. \ref{setup_detail}(b), the setup in PSL is illustrated. Photon pairs that operate at the wavelength of approximately 1550 nm are generated via SPDC using a 15 mm-long type-II periodically poled lithium niobate waveguide (PPLN-WG, HCP), pumped by a 775 nm external-cavity diode laser (ECDL, DL pro 780, Toptica). To stabilize the pump wavelength and the phase of the interferometer, an HWP and a PBS separate part of the power of the pump to a wavemeter (WS6-200 VIS/IR-I, HighFinesse). Using the wavemeter’s built-in locking function, an error signal is generated to provide feedback to the pump laser’s piezo, effectively stabilizing the frequency disturbance of the pump laser to within approximately 1.5 MHz. The final output power of the pump is around 3 mW, and further, it is sent to PPLN-WG via a PM fiber. 

Achieving the HFEQs through HOM interference requires that the photon pairs exhibit degeneracy. To precisely find the degenerate condition, we use the single-photon time-of-flight spectroscopy (ToFS) technique \cite{jin2016simple,davis2017pulsed,chen2017efficient} (Appendix.\ref{SM_ToFS}) to measure the photon pair frequencies and then fine-tune the temperature of the PPLN-WG to achieve degeneracy, confirming that it occurs at 39.5°C with $<$ 10 mK drift (Appendix.\ref{degenerate}).

After photon pair generation, the photon pairs are coupled through a pigtail and passed through a length-adjustable PM fiber to introduce a tunable time delay between orthogonally polarized photons. This mechanism provides a simple setup to control the time delay $\tau_H$ of HOM interference, which also gives a method to manipulate the dimensionality of HFEQs. Before going into the integrated interferometer, the photons are filtered by a long-pass filter (BLP01-980R-25) and a bandpass filter (FBH1550-40, Thorlabs) to remove pump light leakage.

As illustrated in Fig. \ref{setup_th}, Fig. \ref{setup_detail}(b) and discussed in Sec. \ref{secII}, the signal and idler photons were sent into the same unbalanced Michelson interferometer to perform the HOM and Franson interference experiment. This design significantly reduces the technical challenges associated with relative phase fluctuations between photons in separate interferometers while effectively demonstrating the core physics concept of this study.

Despite these advantages, environmental vibrations still slightly affect the system's stability. To mitigate the impact of vibrations on the optical path difference within the interferometer, the entire setup was mounted on an independent, vibration-isolated optical table. Additionally, the interferometer was enclosed within a dual-layer acoustic and thermal insulating shield to minimize phase noise caused by thermal fluctuations and airflow drift.

In the integrated interferometer, a piezo-mounted mirror at the end of the short arm (less than 1 cm) controls the Franson delay, $\tau_F$. The long arm is formed by a reflective light path involving three mirrors. After nine reflections through three mirrors, the free-space path difference of approximately 2.7 meters between the long and short arms is achieved. This significant optical path difference creates a time delay of about 9 ns between photon pair events, enabling effective ToFS analysis of TPES spectral modes (Appendix.\ref{SM_ToFS}) and avoiding single photon interference.

After passing through the Franson interferometer, the photon pairs are separated again using an HWP and PBS, defining two new modes: $s'$-mode and $i'$-mode. These modes are then coupled into optical fibers. The $i'$-mode is directly transmitted via campus optical fiber to the DL, where it is measured by the SNSPD. The SNSPD's output signal is sent to a time controller (ID1000, IDQ), serving as the start trigger for subsequent coincidence counting measurements.

To experimentally demonstrate the spectral features of the HFEQs prepared by the integrated interferometer, the single-mode spectrum of the $s'$-mode photon should be analyzed. First, to suppress sideband effects from its sinc-shaped spectrum (see Appendix.\ref{degenerate}), the photon passes through a Gaussian spectral filter (TOF-1550-SM-L-10-FA, OF-LINK) with a 1.2 nm bandwidth, reshaping it into a Gaussian profile. Single-photon spectral measurements are then conducted using ToFS \cite{chen2017efficient,davis2017pulsed}. The ToFS employs a 12 km dispersion-compensating fiber (AD-SM-C-120-FC/APC-197/197-10) to map frequency information to temporal information via high negative dispersion. Subsequently, the $s'$-mode photon is transmitted through a campus fiber to the DL, where it is detected by the SNSPD. Using $i'$-mode photons as the start trigger, the coincidence measurement between $s'$- and $i'$-mode photons unveils the single-mode spectral characteristics of the $s'$-mode photon. The ToFS achieves a wavelength resolution of approximately 31 pm (see Appendix.\ref{SM_ToFS} for details). Furthermore, the relationship between the Franson delay $\tau_F$ and coincidence counts in TPES (TPI visibility) provides a means to verify the frequency entanglement properties of the HFEQs.


\begin{figure*}[ht]
\centering
\includegraphics[width=0.94\textwidth]{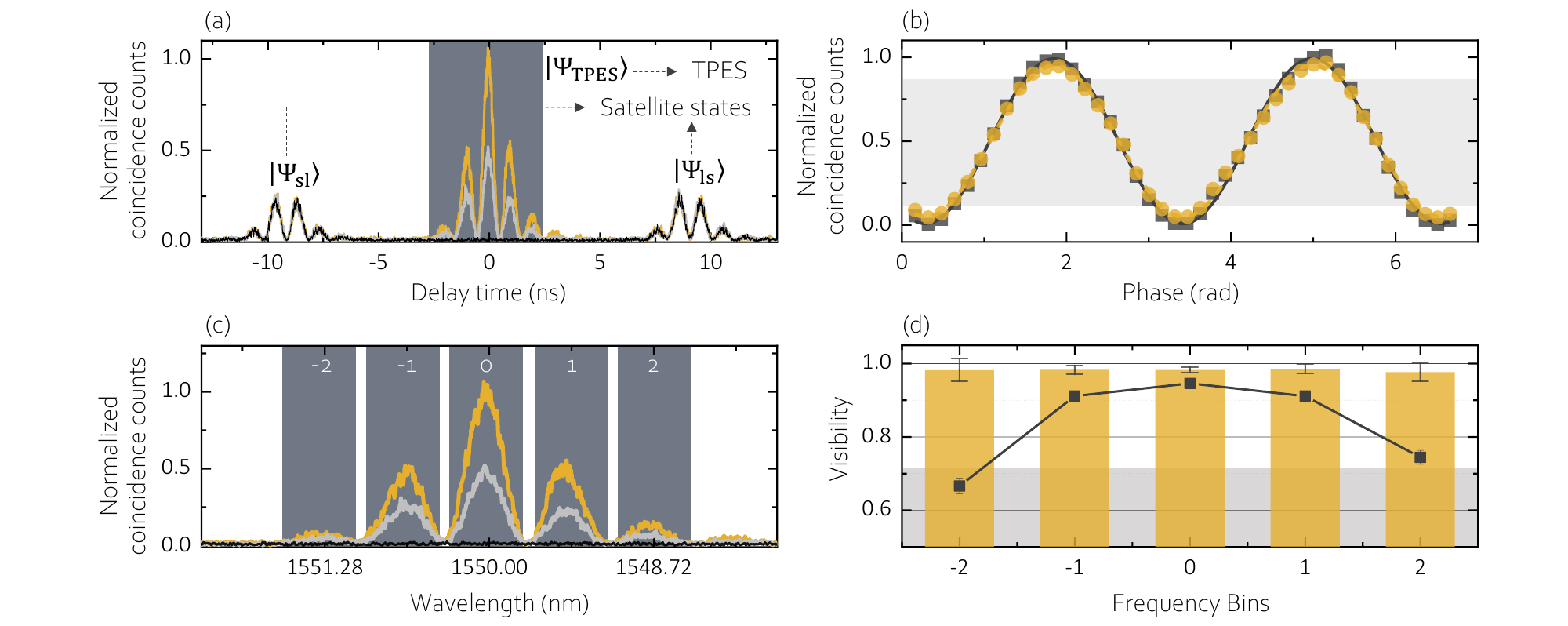}
\caption{(a) Coincidence counting measurement results from the integrated interferometer with a fixed HOM delay ($\tau_H$) and varying Franson delay ($\tau_F$). The coincidence peaks within the blue-gray area represents the TPES. The yellow, gray, and black lines correspond to the constructive, middle, and destructive Franson non-local interference, respectively. (b) TPI fringe. The yellow circles represent the experimental raw data of the coincidence counts, while the black squares denote the data after subtracting the accidental and dark counts. (c) Spectral information of the $s'$-mode of HFEQs with $D=5$. The numbers at the top denote the indices of the frequency bins. (d) Visibilities of each bin of HFEQs. The black squares represent the fitting results from the raw data, while the yellow squares denote the fitting results after subtracting the accidental and dark counts. The error bars, representing one $\sigma$ standard deviation, are estimated based on the Poisson distribution. Appendix. \ref{SM_TPI_FB} provides more details for the TPI fringes for each bin.}
\label{TPI_5bin}
\end{figure*}

\subsection{Experimental results and discussion}
\subsubsection{HFEQs generation and verification}

\begin{figure*}[ht]
\centering
\includegraphics[width=0.94\textwidth]{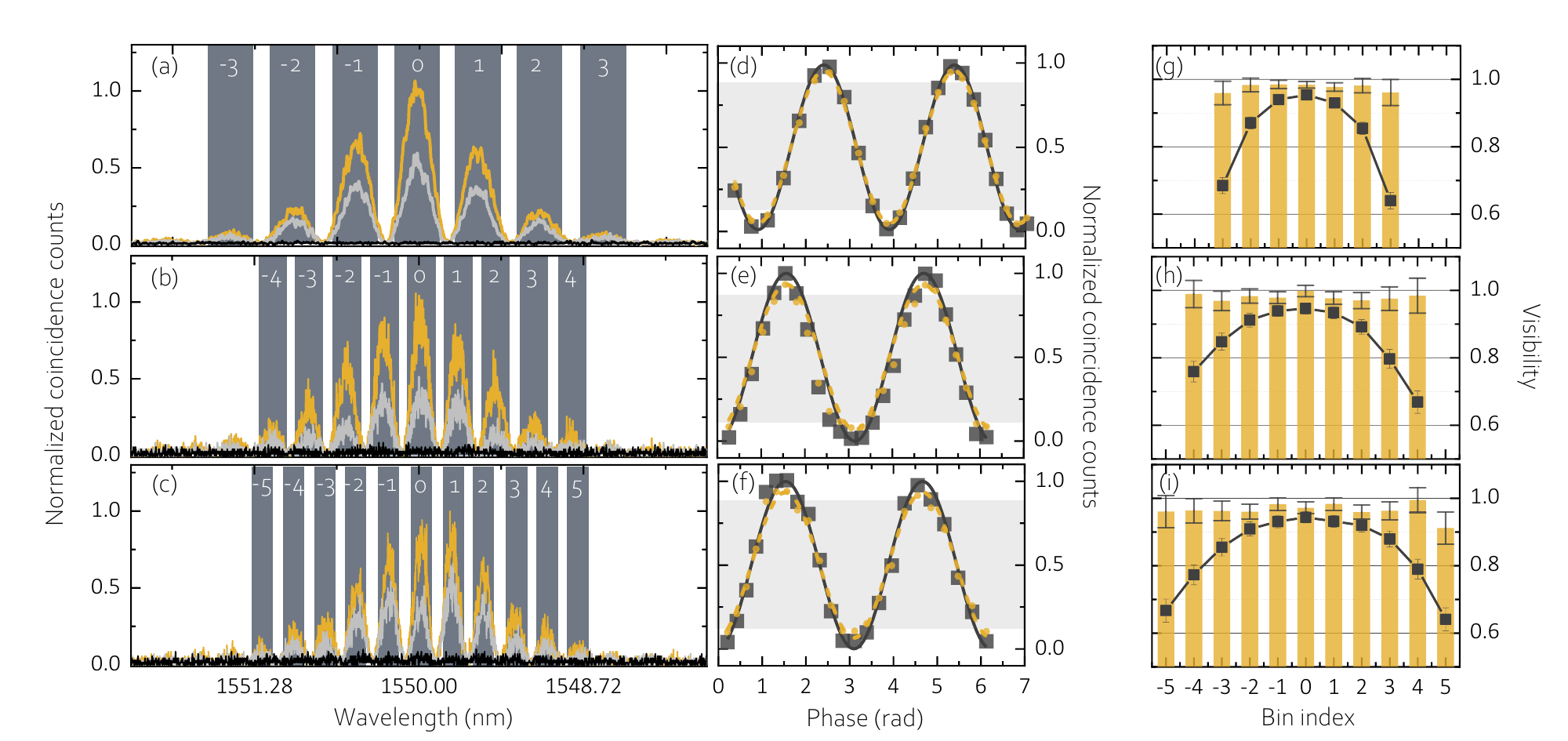}
\caption{Demonstration of dimension manipulation of HFEQs via HOM Delay. (a)-(c) Characterized spectral information of the $s'$-mode of HFEQs with $D = 7, 9,$ and $11$, respectively. The numbers at the top indicate the indices of the frequency bins. The yellow, gray, and black lines correspond to constructive, middle, and destructive Franson nonlocal interference, respectively. (d)-(f) TPI fringes of TPESs. The yellow circles represent the raw experimental coincidence counts, while the black squares denote the data after background subtraction. The black line and yellow dashed line represent the fitting curves for the raw and processed data, respectively. (g)-(i) TPI visibility of each bin in the case of $D=7$, $D=9$, and $D=11$, respectively. The visibility represented by solid squares is estimated from the raw data. The yellow bars show the visibility obtained after subtracting accidental counts. The error bars, showing one $\sigma$ standard deviation, are estimated based on the Poisson distribution. Appendix.\ref{SM_TPI_FB} provides more details for the study of the TPI fringes for each bin. In (e) and (f), the phase is controlled by another method that tunes the pump laser frequency; see Appendix.\ref{phase_ctrl} for details. 
}
\label{TPI_7_11bin}
\end{figure*}

Based on the experimental setup and theoretical proposal described above, we first verify the frequency correlation of the SPDC photon pairs to ensure that their JSI exhibits anti-correlation. The photon pairs are transmitted through a 1-meter PM fiber before entering the Franson interferometer for measurement, with the results shown in Fig. \ref{TPI_5bin}(a). In these measurements, we observe the typical coincidence counting pattern of the Franson interferometer. A group of counts corresponding to the TPES appears near a time difference of 0 ns. Additionally, satellite states associated with the Franson interferometer emerge at approximately $\pm 9$ ns, corresponding to the time delay between the long and short arms, clearly separating the TPES from the satellite states in the time domain, effectively suppressing the single-photon interference.

In Fig. \ref{TPI_5bin}(a), the functionality of ToFS enables the coincidence measurements to reveal additional details about the spectral structure of these quantum states. Notably, both the satellite states and the TPES exhibit a comb-like spectral pattern. This observation suggests that the generated SPDC photon pairs have a non-zero time delay ($\tau_H\neq0$), which induces further spectral modulation via HOM interference. The non-zero $\tau_H$ originates from two primary factors: the 1-meter PM fiber embedded in the waveguide module and the phase difference caused by the refractive index mismatch between the two cross-polarized photon-pair modes within the type-II SPDC chip. Furthermore, we observe that the spectral structures of the satellite states and the TPES result from the modulation effects of destructive and constructive HOM interference, respectively, as predicted by theoretical models.

To verify the frequency correlation characteristics of the TPES, the pizeo at the end of the short arm of the integrated interferometer is scanned to control the Franson delay, $\tau_F$. Simultaneously, the coincidence counts measured for the TPES (the counts within the blue-gray box in Fig. \ref{TPI_5bin}(a)) are calculated and analyzed to observe the TPI fringe, as shown in Fig. \ref{TPI_5bin}(b). The yellow circles in Fig. \ref{TPI_5bin}(b) represent the raw TPI data. Due to the contribution of accidental counts and detector dark counts to the background, the observed TPI visibility is approximately 90.2\%. Nevertheless, the interference contrast significantly exceeds the classical limit of 70.7\% (indicated by the gray region), demonstrating the strong frequency correlation of the SPDC photon pairs. To further illustrate the frequency correlation of the photon pairs, the black squares in Fig. \ref{TPI_5bin}(b) show the coincidence counts after background subtraction—including accidental and dark counts, which are measured within a time window well separated from the TPES and satellite states—to account for experimental imperfections. After fitting, the interference contrast reaches 98.3\%, approaching the theoretically predicted near-perfect nonlocal interference.

After confirming the global frequency correlation characteristics of the TPES, we further analyze its spectral structure. Based on the coincidence counting profile of the TPES shown in Fig. \ref{TPI_5bin}(a), five dominant frequency components are identified ($D=5$). By utilizing the wavelength and photon time-of-flight calibration of the ToFS (Appendix.\ref{SM_ToFS}), we reconstruct the spectral information, as illustrated in Fig. \ref{TPI_5bin}(c). The ToFS reconstruction results reveal that the TPES spectrum exhibits a comb-like structure, modulated by HOM interference. After fitting, the HOM delay $\tau_H$ and mode spacing of $\Delta f=1/2\tau_H$ are estimated at approximately 6.14 ps and 81.43 GHz, respectively, for the TPES with $D = 5$. (Appendix.\ref{fitting}).

Notably, Fig. \ref{TPI_5bin}(c) presents only the single-mode spectrum of the $s'$-mode photons in the TPES. However, for an HFEQ, the frequency correlation between two modes within each frequency bin (i.e., $|J_{A,n}(\Omega_{s,n},\Omega_{i,n})|^2$) should also be verified. To experimentally investigate this, a conventional approach might involve using a spectral filtering device to select the frequency of the $s'$-mode and subsequently performing a TPI measurement. In our system, however, due to the function of ToFS, the spectral information is already encoded in the temporal distribution of coincidence counts. Therefore, by selecting an appropriate time window (in each blue-gray box) within the coincidence data, we can measure the TPI behavior of the individual HFEQs' frequency bins. Based on this concept, we analyze the coincidence counts as a function of $\tau_F$ for individual spectral modes in Fig. \ref{TPI_5bin}(c) to obtain the TPI fringes (see Appendix \ref{SM_TPI_FB}) and their visibilities for each frequency bin, as shown in Fig. \ref{TPI_5bin}(d).

In Fig. \ref{TPI_5bin}(d), the black squares represent the raw visibilities of each frequency bin. It can be observed that the visibilities decrease for frequency bins farther from the central frequency, even dropping below the classical limit of 70.7\%. This decay is primarily due to the background counts, which remain constant in our system. However, the intensity of frequency bins farther from the central frequency follows a Gaussian decay, leading to a corresponding decrease in TPI visibility. To mitigate this issue and highlight the main physical insight, we subtract the background counts, resulting in the visibilities of all bins recovering to approximately 98\%, well above the classical limit of 70.7\%. This result provides strong evidence for the high correlation between the two frequency modes of each bin, i.e., verified $J_{A,n}(\Omega_{s,n},\Omega_{i,n})\neq f_s(\Omega_{s,n})\times f_i(\Omega_{i,n})$. 

Based on this analysis, we confirm the transformation of TPES into an HFEQ with $D=5$, verifying both global and individual frequency correlations. These results strongly support the DV-CV hybrid entanglement structure of HFEQs.

\subsubsection{Manipulation of HFEQs}

In the next, we further demonstrate the manipulation of HFEQ dimensionality through controlling delay times $\tau_H$ in HOM interference.

As illustrated in Fig. \ref{setup_detail}, the time delay in HOM interference is controlled by introducing a phase difference between orthogonally polarized photons using PM fibers of different lengths. To induce different time delays, we sequentially replaced the PM fiber with lengths of 3, 8, and 11 meters. Through ToFS and coincidence counting measurements, we clearly observe that as the PM fiber length increases, the spectral modulation frequency also increases, as shown in Figs. \ref{TPI_7_11bin}(a)-\ref{TPI_7_11bin}(c). This implies the enhancement of the time delay of HOM interference, thereby increasing the dimensionality of the HFEQs.

From the spectral profiles shown in Figs. \ref{TPI_7_11bin}(a)-\ref{TPI_7_11bin}(c), the mode spacings are 60.4, 36.2, and 30.3 GHz, corresponding to $\tau_H=8.27, 13.8,$ and $16.5$ ps, respectively (Appendix \ref{fitting}). Accordingly, the dimensionalities of the HFEQs can be approximately determined as $D = 7, 9,$ and $11$. Furthermore, the frequency correlations of TPESs modulated by different HOM settings were verified through TPI measurements, as shown in Fig. \ref{TPI_7_11bin}(d)-\ref{TPI_7_11bin}(f). Notably, TPESs modulated under different HOM conditions still exhibit clear TPI, with visibilities of $97.6\%$, $99.0\%$, and $97.3\%$, respectively, after subtracting background noise, which demonstrated the high frequency-entanglement of those HFEQs. 

Finally, similar to Fig. \ref{TPI_5bin}(d), we further verify the frequency correlation of individual frequency bins in the prepared HFEQs, with the results presented in Figs. \ref{TPI_7_11bin}(g)-\ref{TPI_7_11bin}(i). As expected, due to existing background counts, the interference visibility of HFEQs decreases for frequency bins farther from the central frequency. However, after subtracting accidental and dark counts, the visibility improves to approximately 95\%-99\%, confirming that the frequency correlations across all frequency bins in the HFEQs for \(D = 7\), 9, and 11 exceed the classical threshold of 70.7\%. This result further validates the strong frequency entanglement present in the generated HFEQs. Furthermore, these results highlight the capability of our setup to control the dimensionality of HFEQs, demonstrating its great potential for HD QIP.

\begin{figure}[t]
\centering
\includegraphics[width=0.45\textwidth]{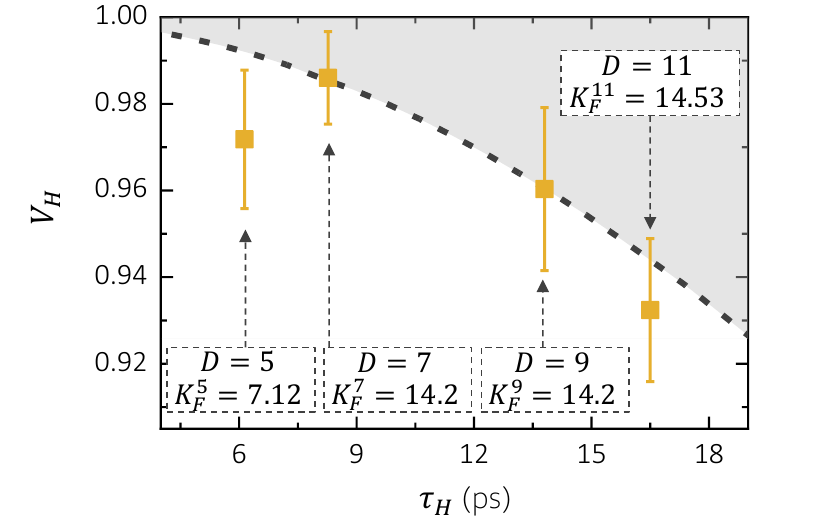}
\caption{HOM visibility as a function of $\tau_H$. The yellow squares represent the experimental data, and the error bars are estimated by fitting the confidence bound. The black dashed line depicts the simulated $V_H$ under a ToFS resolution of 4 GHz ($\approx 31$ pm wavelength resolution). The gray  realm indicates the area that cannot be measured due to experimental limitations.
}
\label{visibility_tau}
\end{figure}

\subsubsection{Quantification of entanglement of HFEQs}

Based on the above discussion and tests, the frequency structure, dimensionality, and individual and global frequency correlations of the HFEQs have been fully characterized using HOM and Franson interference. The obtained high visibility of Franson interference indicates that the generated HFEQs exhibit strong frequency anti-correlations, providing confidence in estimating their entanglement strength, the Schmidt number $K_F$, using the approach proposed in Sec. \ref{thoeory_HOM}. To achieve this, the HOM visibilities of each HFEQ in Fig. \ref{TPI_5bin}(c) and Figs. \ref{TPI_7_11bin}(a)-\ref{TPI_7_11bin}(c) are individually analyzed using the method discussed  in Appendix.\ref{fitting} with respect to their corresponding $\tau_H$. The results are presented by the yellow squares in Fig. \ref{visibility_tau}. In this result, we can observe a general trend: $V_H$ begins to decrease as $\tau_H$ increases. The underlying causes of this trend, as discussed in Sec. \ref{thoeory_HOM}, may stem from two primary factors. On the one hand, it could be due to the finite pump bandwidth or dephasing that causes the reduction of entanglement. On the other hand, the finite frequency resolution of the ToFS may also contribute to the observation of a reduced $V_H$.

To understand the primary cause of the decrease in $V_H$, the black dashed line in Fig. \ref{visibility_tau} represents the relationship between HOM interference visibility and $\tau_H$, when taking into account the spectral blurring due to the finite frequency resolution of ToFS (Appendix.\ref{jitter_app}). This trend shows that as $\tau_H$ increases and the HOM modulation frequency rises, the finite resolution (31 pm) of ToFS limits a complete resolution of the fine structure of the single-mode spectrum, leading to a reduction in $V_H$. On the other hand, by comparing the experimental results with the simulations, we observe that the measured HOM interference visibility of the HFEQs essentially reaches the upper limit imposed by the experimental apparatus.

Based on the measurement results in Fig. \ref{visibility_tau}, we conclude that the reduction of $V_H$ is primarily due to the finite resolution of ToFS. This result validates the reliability of using the fitted $V_H$ and $\tau_H$ to determine the lower bound of the corresponding Schmidt number $K_F$. By referencing the relationship in Fig. \ref{HOM_visibility}, we obtain the Schmidt numbers for the generated HFEQs with dimensions $D=5, 7, 9$, and $11$ as $K_F=7.12, 14.2, 14.2$, and $14.53$, respectively. All values exceed the maximum Schmidt number of conventional FEQs with the same dimensionality, confirming the stronger frequency entanglement of the generated HFEQs.

More importantly, the high-contrast comb-like spectral structure observed in Figs. \ref{TPI_7_11bin} and \ref{visibility_tau} strongly indicates that the generated HFEQs are pure states. This implies that all observed frequency correlations stem from genuine frequency entanglement.

\section{Summary}\label{secIV}

In summary, this work presents a comprehensive theoretical and experimental investigation of an integrated quantum interferometer. Building on this foundation, we further explore the novel quantum states of HFEQs.

In theory, we investigate an integrated interferometer that incorporates two types of quantum interference: HOM interference and Franson-type TPI. These interference effects allow modulation of the joint spectral structure along the sum-frequency and difference-frequency axes, respectively. Under the influence of HOM interference, a strongly anti-correlated photon pair can be effectively transformed into a hybrid CV-DV quantum state, referred to as an HFEQ.

Experimentally, we implemented a system to demonstrate the generation, manipulation, verification, and distribution of HFEQs over a 770 m commercial optical fiber link spanning two campus buildings. Utilizing a simple tunable HOM interference mechanism, we employed ToFS to confirm the DV nature of HFEQs and demonstrated control over their dimensions.

Furthermore, by performing Franson TPI measurements on individual frequency bins of HFEQs, we confirmed that the TPI contrast for all frequency bins remained between 95\% and 98\%, significantly exceeding the classical limit of 70.7\%. This result clearly manifests the highly entangled CV nature in each bin of the HFEQs. 

Moreover, based on these results, we estimated the Schmidt number of HFEQs using our theoretical method. These values surpass the maximum Schmidt numbers achievable by conventional FEQs, further confirming the superior frequency entanglement properties of HFEQs. Table \ref{tab:HFEQ} summarizes the key experimental results of the prepared HFEQs.

\begin{table}[h]
    \centering
    \caption{Summary of the performance of HFEQs. $V_F$ is the TPI visibility of HFEQs after subtracting the background counts, and the value in the parentheses represents the raw visibility. $K_F^{D}$ denotes the lower bound of the Schmidt number of HFEQs with dimension $D$. }
    \begin{tabular}{|c|c|c|c|c|}
    \hline
     $D$   & $V_H(\%)$ & $V_F(\%)$ & Mode spacing, $1/2\tau_H$ (GHz) & $K_F^{D}$  \\ \hline
     5 & 97.2  & 98.3(90.2) & 81.43 & 7.12 \\ \hline
     7 & 98.86 & 97.6(89.1) & 60.4 & 14.2  \\ \hline
     9 & 96.03 & 99.0(90.4) & 36.2 & 14.2  \\ \hline
     11& 93.24 & 97.3(85.7) & 30.3 & 14.53 \\ \hline
    \end{tabular}
    \label{tab:HFEQ}
\end{table}

\section{Conclusion}\label{secV}

In conclusion, this study uncovers, to the best of our knowledge, several previously unexplored physical phenomena, demonstrating significant potential for applications in HD time-frequency QIP. 

First, by analyzing the proposed integrated interferometer, we have shown its capability for full control and verification of time-frequency quantum information in photon pairs, thus enabling the generation of meaningful quantum states in this domain. Notable examples include time-frequency grid states (discussed in Appendix \ref{GKP_states}) for quantum error correction \cite{descamps2024gottesman} and the HFEQs investigated in this study.

More importantly, this work reveals deeper physical insights into HOM-based FEQs. The results demonstrate that after undergoing HOM interference, SPDC photon pairs with high frequency-anti-correlations still exhibit strong correlations and entanglement within each discrete frequency bin. This implies that FEQs established using HOM interference cannot always be described by $|\Psi_{\text{FEQ}}\rangle$ in Eq. (\ref{FEQs_state}), which represents a mere superposition of several correlated discrete frequencies. Instead, they should be more accurately represented by $|\Psi_{\text{HFEQ}}\rangle$ in Eq. (\ref{HFEQs_state}), which accounts also for the CV entanglement within each frequency bin. 

The fundamental distinction between $|\Psi_{\text{FEQ}}\rangle$ and $|\Psi_{\text{HFEQ}}\rangle$ lies in their internal structure: for FEQs, each basis state $|\omega_n, \omega_{D-n}\rangle$ can be independently decomposed into two pure states, i.e., $|\omega_n\rangle\otimes|\omega_{D-n}\rangle$. However, in HFEQs, each frequency bin forms a CV entangled state, preventing such a decomposition, i.e.,
\begin{equation}
\begin{aligned}
\int\int d\Omega_sd\Omega_iJ_{A,n}(\Omega_{s,n},\Omega_{i,n})|\Omega_{s},\Omega_{i}\rangle\\
\neq\int d\Omega_sf_{s,n}(\Omega_s)|\Omega_{s}\rangle\otimes\int d\Omega_if_{i,n}(\Omega_i)|\Omega_{i}\rangle.
\end{aligned}
\end{equation}
Therefore, tracing out one frequency mode results in a mixed state rather than a pure state. This effect may pose a critical issue when utilizing HOM-based FEQs as fundamental resources for subsequent time-frequency QIP. 

From this perspective, an intriguing question arises: is there an optimal initial SPDC JSA that minimizes the entanglement in each frequency mode, thereby producing pure FEQs? Or, how can one choose an optimal HOM condition for generating pure FEQs? This question represents an important avenue for future research, aimed at further optimizing the generation of HOM-based FEQs and exploring their fundamental limits in time-frequency QIP.

On the other hand, while the entanglement within each frequency mode might be considered a limitation for FEQ applications, it could potentially enable novel applications for HFEQs. For instance, in quantum key distribution (QKD), the discrete frequency bins allow communication parties to use wavelength-division multiplexing (WDM) to separate each frequency bin and couple them into a single optical fiber, thereby enhancing data transmission rates and reducing cross-talk errors \cite{kovalenko2021frequency}. Simultaneously, the CV quantum information (frequency entanglement) carried within each frequency bin can facilitate additional time-frequency HD QKD \cite{nunn2013large,zhang2014unconditional}, even forming a hybrid DV–CV time-frequency QIP system. Additionally, in CV quantum information-based quantum sensing, such quantum states may offer a useful resource as well \cite{cardoso2021superposition}.

This work establishes a novel resource for HD frequency-encoded QIP, providing not only a deeper theoretical understanding but also a practical foundation for future quantum technologies.

\begin{figure*}[t]
\centering
\includegraphics[width=0.95\textwidth]{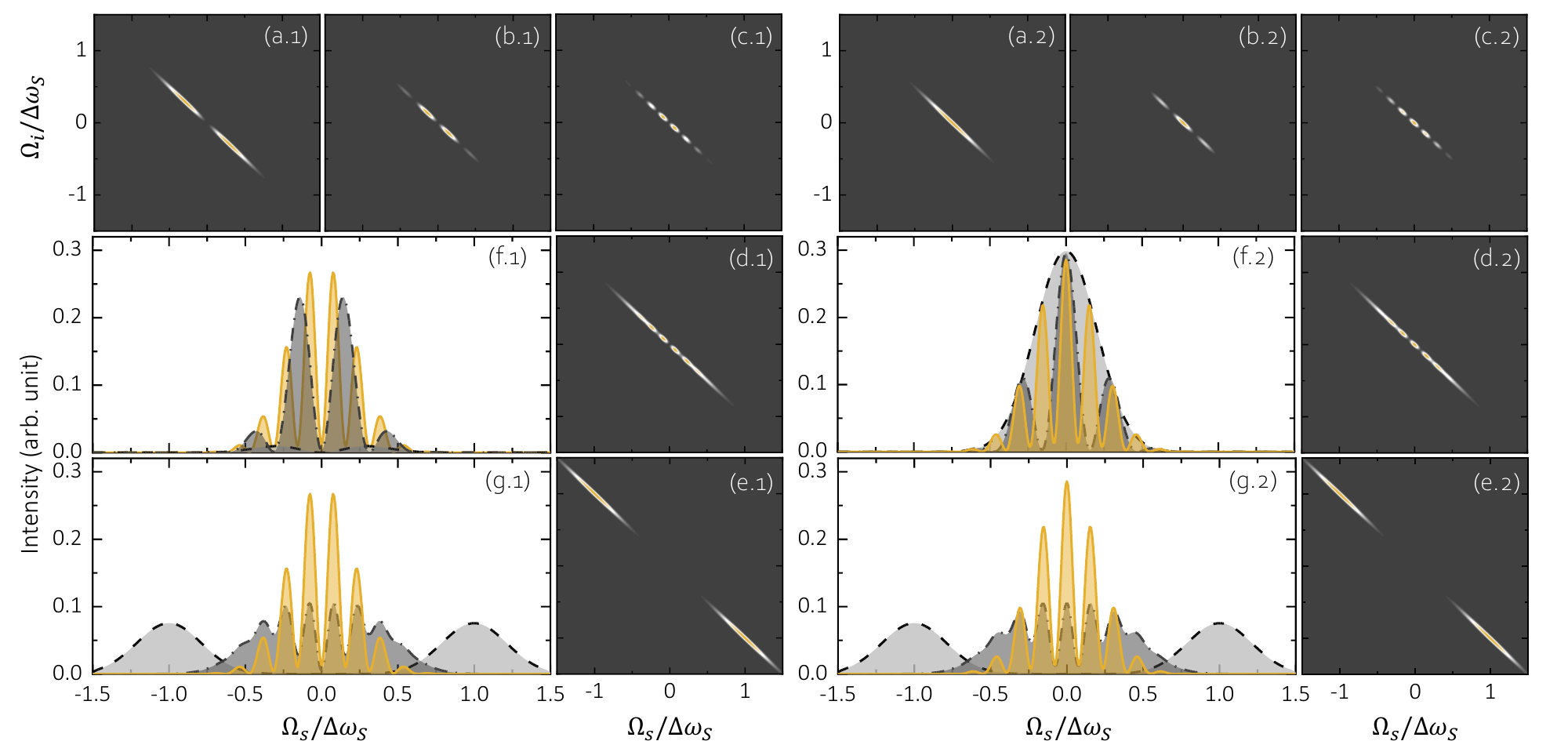}
\caption{The demonstration illustrates a series of SPDC JSIs under (a.1)-(g.1) destructive and (a.2)-(g.2) constructive HOM interference. In (a.1)-(c.1) and (a.2)-(c.2), we examine a degenerate JSI with a series of HOM delay times $\tau_H$ of $1/\Delta\omega_S$, $10/\Delta\omega_S$, and $20/\Delta\omega_S$, respectively. (c.1)-(e.1) and (c.2)-(e.2) show the evolution of the HOM JSI as it transitions from the degenerate to the non-degenerate condition. (f.1) and (f.2) show the reduced signal mode spectrum of (a.1)-(c.1) and (a.2)-(c.2), respectively. Notably, in the condition of (a.1) that is shown in (f.1), the single-photon spectrum reflects the impact of destructive HOM interference when $\tau_H$ is small. This leads to a significant reduction in spectral intensity. (g.1) and (g.2) show the reduced signal mode spectrum of (c.1)-(e.1) and (c.2)-(e.2), respectively. $\Delta\omega_p$ is set to $\Delta\omega_S/20$ for all cases.}
\label{JSI_series1}
\end{figure*}

\section*{Acknowledgments}
This work was supported by the National Science and Technology Council (NSTC) of Taiwan under Grants 112-2112-M-008-025-MY3, 113-2119-M-008-010, 113-2923-E-008-001, and 114-2927-I-008-501. We also thank Dr. Kuan-Ting Lin at the National Taiwan University and Dr. Bo-Han Wu at the Massachusetts Institute of Technology for their fruitful discussions. Especially thank Dr. Ying-Cheng Chen at the Institute of Atomic and Molecular Sciences, Academia Sinica, for the instruments support. 

S.-H. W. and P.-H. C. contributed equally to this work.

\section*{Appendix}
\appendix
\section{Discussion on the satellite states and the effect of degeneracy}\label{satellite_states}

In this section, we complement the discussion on satellite states which also shows some fascinating physics here. To demonstrate that, we analyze the JSI of satellite states $|\Psi_{\text{ls}}\rangle$ and $|\Psi_{\text{sl}}\rangle$ in Eq. (\ref{phi_4}). Their JSI $J_I^{ls}$ and $J_I^{sl}$ is given by 
\begin{equation}
\begin{aligned}
J^{ls,sl}_I(\Omega_s,\Omega_i)&=\left|\langle\Omega_s,\Omega_i|\Psi_{\text{ls,sl}}\rangle\right|^2\\
&=\frac{1}{16}|f(\Omega_s,\Omega_i)-f(\Omega_i,\Omega_s)e^{-i(\Omega_s-\Omega_i)\tau_H}|^2,
\label{JSI_sl}
\end{aligned}
\end{equation} 
In Eq. (\ref{JSI_sl}), we observe that the typical HOM effect causes destructive interference along the difference frequency, $\Omega_-$, between two SPDC JSAs, each of which has exchanged frequency components and a phase difference resulting from the previously introduced time delay, $\tau_H$. Notice that this is different from the TPES, which shows constructive HOM interference. 

Destructive HOM interference arises from the path length difference of satellite states. Due to this path length difference, satellite states do not interfere at PBS2 in Fig. \ref{setup_th}. However, at PBS1, satellite states correspond to photons with different polarization inputs, leading to their destructive HOM interference \cite{kim2017two}.

In Eq. (\ref{JSI_sl}), that is clear to see that, the maximum HOM effect will happen in the special case where the degenerate SPDC JSA has a symmetric structure, such that $f(\Omega_s, \Omega_i) = f(\Omega_i, \Omega_s)$, and Eq. (\ref{JSI_sl}) can be further simplified and expressed as
\begin{equation}
\begin{aligned}
J^{ls,sl}_I(\Omega_s,\Omega_i)&=\frac{1}{8}|f(\Omega_s,\Omega_i)|^2\left[1-\cos(\Omega_-\tau_H)\right].
\label{JSI_sl2}
\end{aligned}
\end{equation}
A notable insight can be drawn from Eq. (\ref{JSI_sl2}). For $\tau_H=0$, total destructive interference occurs, resulting in zero probability of detecting the states $|\Psi_{\text{sl}}\rangle$ and $|\Psi_{\text{ls}}\rangle$. This behavior differs significantly from that of a conventional Franson interferometer, where $|\Psi_{\text{sl}}\rangle$ and $|\Psi_{\text{ls}}\rangle$ always exist. However, this phenomenon has neither been thoroughly discussed nor observed in previous studies with similar setups \cite{park2018time}.  

One possible reason is that the SPDC process in these studies often involved highly non-degenerate photon pairs. Under this condition, the SPDC JSAs in Eq. (\ref{JSI_sl}) become two entirely different functions when frequency-exchange mechanisms are applied, preventing complete destructive interference even when $\tau_H = 0$. Moreover, even in the degenerate SPDC case, complete destructive interference remains unattainable when $\tau_H \neq 0$. Therefore, achieving total destructive HOM interference for satellite states requires a strict set of conditions to be met.

Even so, in the near-degenerate case, the original JSI is modulated by difference-frequency interference. Notably, from the perspective of time-frequency quantum information, this modulation process transforms $|\Psi_{\text{sl}}\rangle$ and $|\Psi_{\text{ls}}\rangle$ into a correlated, comb-like structure in the joint spectral domain, thereby enabling this optical setup to effectively generate an HFEQ.

In Fig. \ref{JSI_series1}(a.1)-\ref{JSI_series1}(g.1), we illustrate the behavior of $J^{ls,sl}_I(\Omega_s, \Omega_i)$ for both degenerate and non-degenerate cases across a range of delay times, $\tau_H$. In the highly non-degenerate case, as shown in Fig. \ref{JSI_series1}(e.1), the HOM interference has little to no observable effect. In contrast, as SPDC approaches degenerate (Fig. \ref{JSI_series1}(c.1)-\ref{JSI_series1}(e.1)), the HOM effect becomes more pronounced, further modulating the JSI into a comb-like structure and gradually forming an HFEQ. Additionally, increasing the time delay $\tau_H$, as seen in Fig. \ref{JSI_series1}(a.1)-\ref{JSI_series1}(c.1), leads to an increase in the frequency of HOM interference fringes, resulting in a higher-dimensional HFEQ. In Fig. \ref{JSI_series1}(a.2)-\ref{JSI_series1}(g.2), we also provide a demonstration of the constructive HOM interference in $J^{\text{TPES}}_I(\Omega_s, \Omega_i)$ for both degenerate and non-degenerate cases for comparison.

\section{Time-frequency grid state}\label{GKP_states}

In Fig. \ref{JSI_series2}(f) and \ref{JSI_series2}(i), an intriguing phenomenon is observed when the JSI exhibits weaker correlations, specifically when $\tau_F = \tau_H$. In those cases, the interference frequencies of the sum frequency and difference frequency align, resulting in a grid-like JSI distribution for the TPES. This feature is especially pronounced in the conditions illustrated in Fig. \ref{JSI_series2}(i). This observation underscores the emergence of a novel type of Gottesman-Kitaev-Preskill state in the time-frequency domain, with potential applications in quantum error correction protocols utilizing Bosonic rotation codes for time-frequency quantum information \cite{yamazaki2023linear,grimsmo2020quantum}.

Importantly, Eq. (\ref{JSA_llss2}) describes the JSA of the TPES $|\Psi_{\text{TPES}}\rangle$ after conditional filtering, which excludes the satellite states $|\Psi_{\text{sl}}\rangle$ and $|\Psi_{\text{ls}}\rangle$. In this case, the interference directions align with the difference-frequency and sum-frequency axes between the signal and idler photons. This result contrasts with the interference patterns observed in the unfiltered folded Franson interference JSI output, where the directions are purely along the signal and idler frequencies \cite{jin2024spectrally}. The nature of this integrated interferometer provides additional potential applications.


\begin{figure}[ht]
\centering
\includegraphics[width=0.45\textwidth]{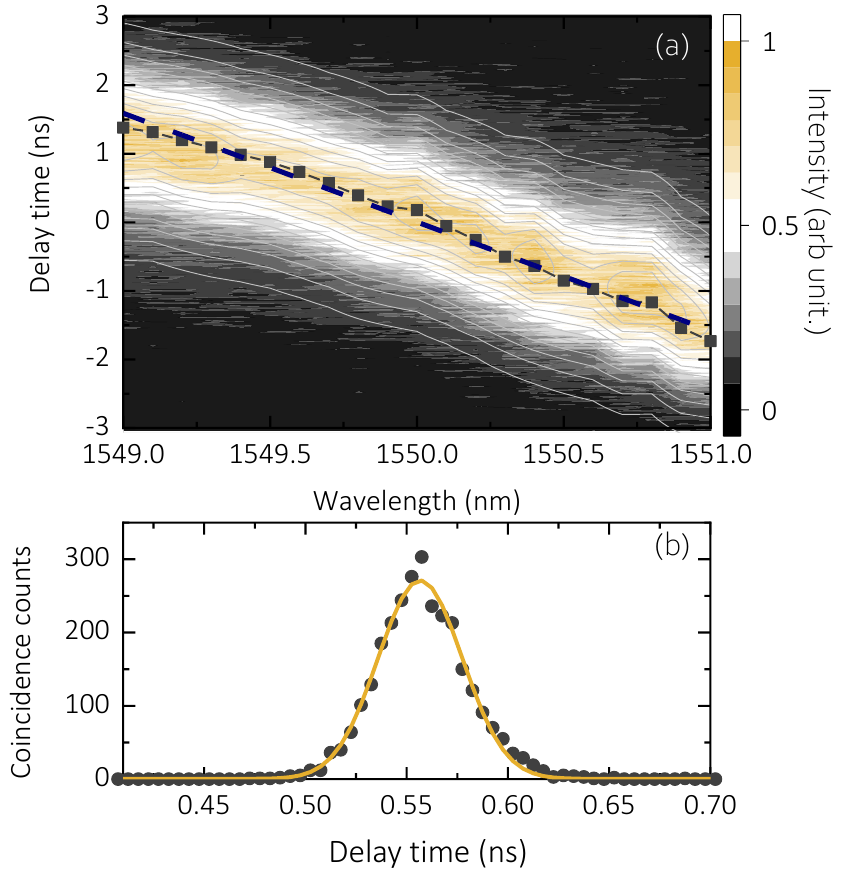}
\caption{The relationship between the wavelength and time delays of ToFS. The gray squares denote the experimental data, and the solid dashed blue line represents the linear fit given by $t=-1.58597\lambda_s+2458.26$, where $t$ is the delay time. The background color map illustrates the coincidence count distribution at different central wavelengths of the filter. (b) Testing the timing jitter of the experimental system. The black dots indicate the experimental data of coincidence counts of photon pairs without DCF, while the yellow line represents the Gaussian fit of the distribution with an FWHM of 49.1 ps.}
\label{tofs}
\end{figure}

\section{Detail of ToFS}\label{SM_ToFS}

The spectral measurement of single photons (or photon pairs) is accomplished using a time-of-flight spectrometer (ToFS). The operation of the ToFS is based on the frequency-to-time mapping technique, which employs a dispersive medium to convert the frequency information of a single photon into its time of flight, enabling the reconstruction of the single-photon spectrum \cite{chen2017efficient,davis2017pulsed}. The key component of the dispersive medium in our ToFS is constructed using a 12 km dispersion-compensating fiber (DCF), which provides a dispersion rate of -2.014 ns/nm @1545 nm and has an intrinsic loss of 7.33 dB. Compared to a standard single-mode fiber, achieving the same dispersion rate would require approximately 143.8 km in length and would exhibit a relatively high loss of 28.76 dB. These comparisons highlight the advantages of this component as a ToFS.

To precisely convert the single-photon time of flight into wavelength, we employed the following approach. First, all HWPs in Fig. \ref{setup_detail}(b) are set to 0 degrees to prevent frequency exchange between the signal and idler photons, ensuring that they travel along the long and short arms, respectively, before being separated at PBS2.

The photon in the $i'$-mode is used as the trigger signal for the coincidence measurement to establish the time reference, while the photon in the $s'$-mode passes through the DCF, is detected, and its time of flight is recorded. To determine the relationship between photon wavelength and time of flight, a tunable fiber filter (bandwidth 1.2 nm) is placed in the $s'$-mode to selectively transmit photons at specific wavelengths. Additionally, the crystal temperature is slightly adjusted to ensure sufficient photon transmission through the tunable fiber filter, maintaining a high signal-to-noise ratio. 

By recording the central transmission wavelength set by the tunable fiber filter and the time of flight of the $s'$-mode photon, we established their relationship to calibrate the ToFS. The results are shown in Fig. \ref{tofs}(a). The color map represents the coincidence count distribution at different central wavelengths of the filter. To determine the delay time of each coincidence count distribution, Gaussian pulse fitting is applied, as shown by the white solid line. Consequently, the relationship between the signal photons' wavelength and delay time is obtained. After performing a linear fit, the resulting equation is given by $t = -1.58597\lambda_s + 2458.26$, where $t$ represents the time of flight (in ns) and $\lambda_s$ is the photon wavelength (in nm). This equation already accounts for time differences caused by disparities in the optical path lengths of the $s'$-mode and $i'$-mode in the experimental setup.

The fitting results indicate that the ToFS provides a dispersion rate of $-1.58597$ ns/nm within the entire system. This value is slightly lower than the specified dispersion of the DCF used in our setup, likely due to the presence of multiple fiber networks in the system, including campus optical fibers and building-interior fiber connections.

Furthermore, based on the measured dispersion rate, we can infer the frequency resolution of the ToFS. The resolution limit of the ToFS is primarily constrained by the jitter in detecting photon time of flight. To determine the precise timing jitter, we temporarily removed the DCF from the experimental setup and performed a coincidence count measurement of photon pairs to characterize the overall system timing jitter, as shown in Fig. \ref{tofs}(b). According to the results, the timing jitter of the experimental system is approximately 49.1 ps. The main contributions to this jitter originate from the SNSPD (18 ps rms) and the time controller’s two channels (each contributing 4 ps rms). When converted to a full-width at half-maximum (FWHM) of the time jitter, the estimated value is approximately 44.4 ps, which aligns well with the experimental measurement. In conjunction with the ToFS dispersion rate, the wavelength resolution is approximately $31$ pm, corresponding to a frequency resolution of $\Delta\omega_j = 4$ GHz.

\section{Fitting function for HOM-modulated single-mode spectrum}\label{fitting}

To obtain more precise information about the TPES under the HOM modulation, we consider Eq. (\ref{SPS}) and Eq. (\ref{JSA_llss2}) to further fit the results in Fig. \ref{TPI_5bin}(c) and Fig. \ref{TPI_7_11bin}. However, in most of the cases, Eq. (\ref{SPS}) does not have a simple analytical form. To derive a simplified function for describing the single-mode spectrum of HOM-modulated TPES, we first ignore the Franson effect in Eq. (\ref{JSA_llss2}) to form the JSI and consider an extreme condition where the pump bandwidth is very narrow, in which $\Delta\omega_p\rightarrow0$. Under this assumption, the pump spectrum can be seen as a $\delta$-function, and thus, Eq. (\ref{SPS}) can be calculated, 
\begin{equation}
\begin{aligned}
S_s(\Omega_s,\tau_H)\propto e^{-4\ln{2}\frac{4\Omega_s^2}{\Delta\omega_S^2}}[1+\cos(2\Omega_s\tau_H)].
\label{fitting_eq}
\end{aligned}
\end{equation}
It is particularly noteworthy that Eq. (\ref{fitting_eq}) describes a single-mode spectrum corresponding to an SPDC JSI with perfect frequency correlations. However, as observed in Fig. \ref{HOM_visibility}, a finite pump bandwidth leads to a reduction in HOM visibility. Furthermore, in practical experiments, the finite frequency resolution of ToFS also contributes to the reduction of interference contrast. To account for these effects, Eq. (\ref{fitting_eq}) is phenomenologically modified to incorporate the influence of imperfect contrast, as
\begin{equation}
\begin{aligned}
S_s(\Omega_s,\tau_H)=\frac{N_0}{2}e^{-4\ln{2}\frac{4\Omega_s^2}{\Delta\omega_S^2}}[1+V_H\cos(2\Omega_s\tau_H)],
\label{fitting_eq1}
\end{aligned}
\end{equation}
where $N_0$ is the peak count of the single-mode spectrum, and $V_H$ is the fitted HOM visibility. In addition, the HOM delay $\tau_H$, the mode spacing of $\Delta f=1/2\tau_H$, and the bandwidth of HFEQs, $\Delta\omega_S$, can also be obtained using this equation.

\section{Simulating the effect of ToFS finite resolution}\label{jitter_app}

Experimentally, the single-photon spectral measurement of ToFS is limited by a certain resolution due to disturbances in estimating the photon time of flight. These disturbances arise from factors such as detector timing jitter and counter-timing jitter, introducing a finite frequency resolution into the measurement process. To account for ToFS's finite resolution, we model the frequency measurement jitter using a Gaussian probability distribution,
\begin{equation}
\begin{aligned}
P_{j}(\omega,\Omega_s)=\sqrt{\frac{4\ln2}{\pi\Delta\omega_j^2}}e^{-4\ln{2}\frac{(\omega-\Omega_s)^2}{\Delta\omega_j^2}},
\label{jitter}
\end{aligned}
\end{equation}
where $\Delta\omega_j$ represents the frequency resolution of ToFS, which defines the linewidth broadening in measurement. The function $P_{j}(\omega,\Omega_s)$ describes the probability distribution of a photon with frequency $\Omega_s$ in the ToFS measurement. For instance, if a photon carries a frequency of $\Omega_s$, the finite resolution of ToFS results in a Gaussian-distributed spectral measurement centered at $\Omega_s$.

To extend this to multimode photon frequency measurements, we model all photon frequencies as passing through a Gaussian response function with finite frequency resolution. This means that the final measured spectrum is obtained by convolving the original spectrum with this Gaussian function. Mathematically, the measured spectrum can be expressed as
\begin{equation}
\begin{aligned}
S_j(\omega,\tau_H)=\int d\Omega_s S(\Omega_s,\tau_H)P_{j}(\omega,\Omega_s).
\label{jitter2}
\end{aligned}
\end{equation}
This convolution process effectively models the spectral broadening caused by finite resolution. Using Eq.(\ref{jitter}), Eq. (\ref{jitter2}), and Eq. (\ref{fitting_eq1}), with a frequency resolution of $\Delta\omega_j = 4$ GHz and varying $\tau_H$, the upper limit of the measured $V_H$ is calculated using Eq. (\ref{V_H}), as shown in Fig. \ref{visibility_tau}.

\section{Photon pair degenerate}\label{degenerate}

\begin{figure}[t]
\centering
\includegraphics[width=0.45\textwidth]{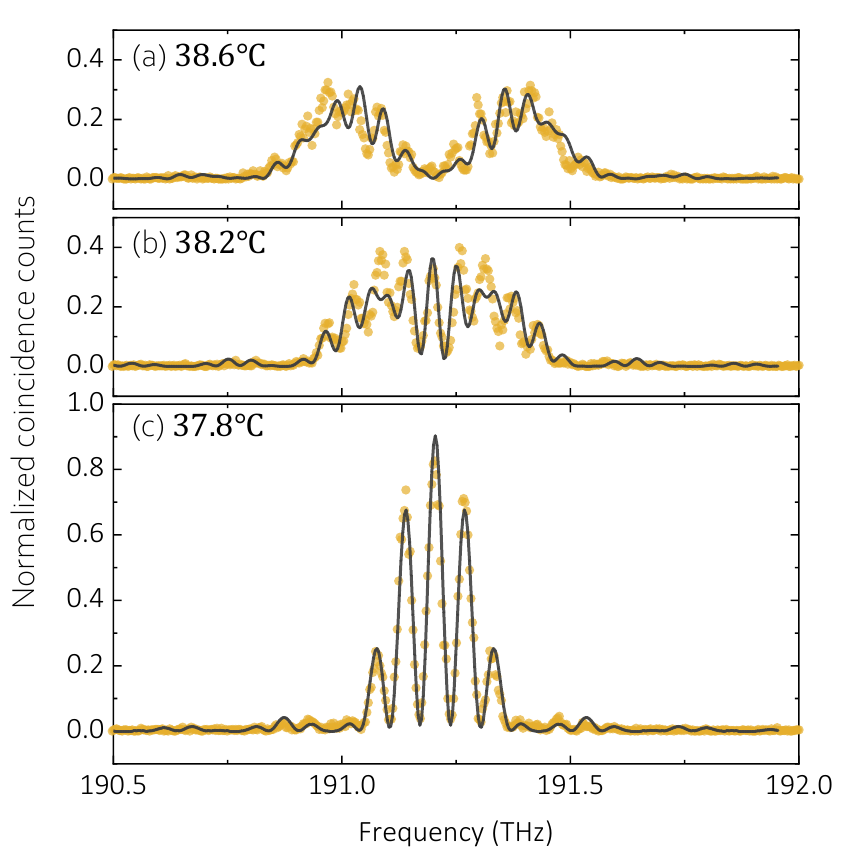}
\caption{Demonstration of tuning the photon pair to degeneracy and the transition of HOM interference. The labels in (a)-(c) indicate the temperature of the crystal, which also corresponds to the degree of degeneracy of the photon pairs. The yellow dots represent the experimental data, and the black line is the fitting results based on the function of
$S_s(\Omega_s,\tau_H)\propto\int d\Omega_i\left|f(\Omega_s,\Omega_i)+f(\Omega_i,\Omega_s)e^{-i\Omega_{-}\tau_H}\right|^2$, where $f(\Omega_i,\Omega_s)=\sinc(\Omega_-/\Delta\omega_S)e^{-2ln(2)\Omega_{+}^2/\Delta\omega_p^2}$.
}
\label{photon_degenerate}
\end{figure}

\begin{figure*}[ht]
\centering
\includegraphics[width=0.95\textwidth]{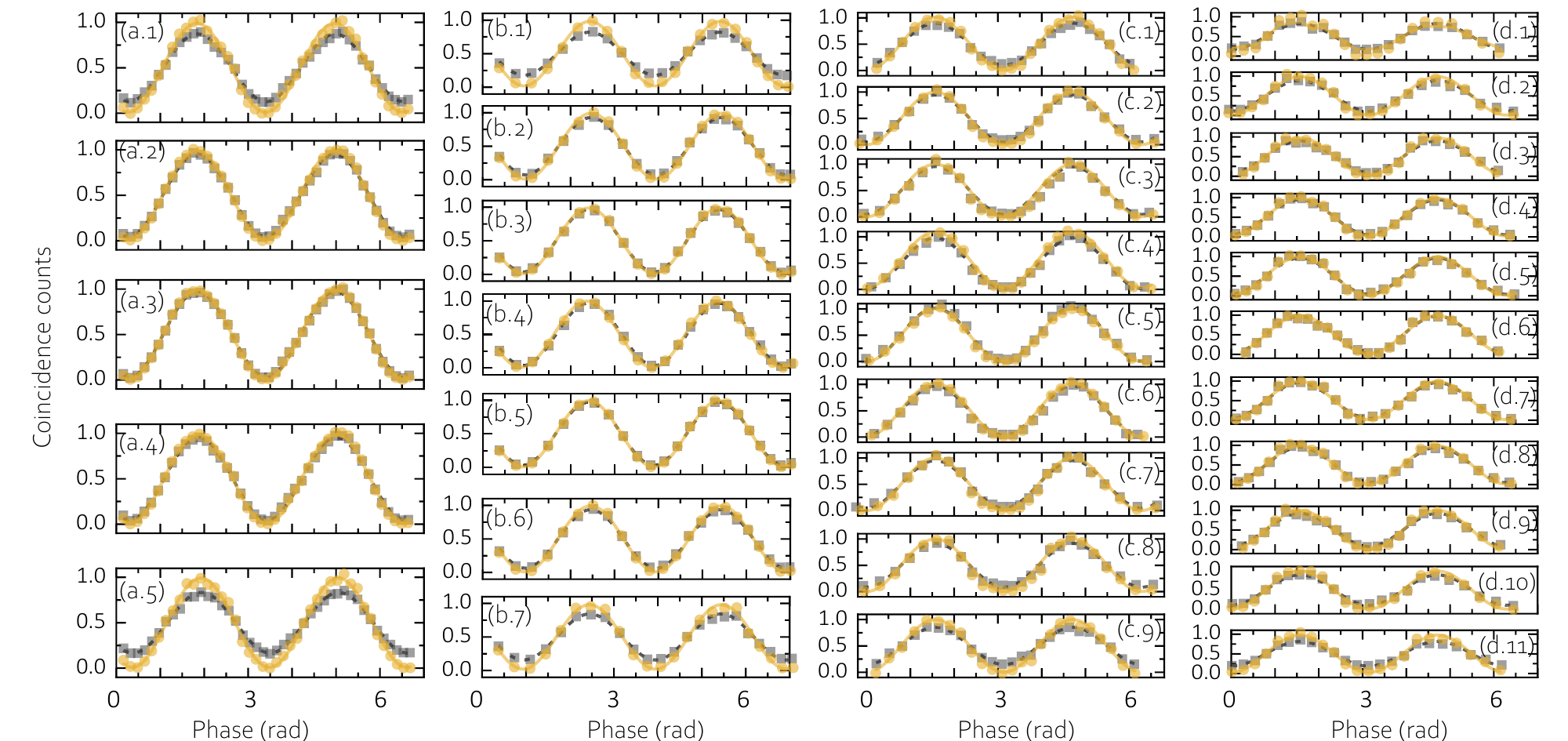}
\caption{TPI fringes for each frequency bin of the HFEQs in the dimensionality of (a.1)-(a.5) $D=5$, (b.1)-(b.7) $D=7$, (c.1)-(c.9) $D=9$, and (d.1)-(d.11) $D=11$. The bin index of each figure decreases from top to bottom. The black squares represent the experimental raw data of the coincidence counts, while the yellow circles denote the data after subtracting the accidental and dark counts. The black and yellow lines represent the theoretical fitting. }
\label{TPI_FB}
\end{figure*}

The precise degeneracy of the photon pair wavelengths is crucial for generating an HFEQ. In this section, we demonstrate the method used in our system to verify photon-pair degeneracy. Additionally, this demonstration provides insights into the gradual transition of photon-pair degeneracy and its relationship with HOM interference.

In Fig. \ref{JSI_series1}(c.2)-\ref{JSI_series1}(e.2), we calculate the JSI of TPES in the non-degenerate regime using Eq. (\ref{JSA_llss}), excluding the Franson interference. Additionally, Fig. \ref{JSI_series1}(g.2) presents the corresponding single-mode spectrum. We observe that as the photon pair transitions from a non-degenerate to a degenerate state, the two main spectral distributions, induced by the frequency exchange mechanism, gradually converge. Once they overlap, HOM interference occurs. Thus, by examining whether the frequency distribution of the $s'$-mode single-mode spectrum consists of a single dominant peak, we can determine whether the photon pair is in a degenerate state. 

To verify whether the photon pair is in a degenerate state in the experiment, the long arm of the Franson interferometer in Fig. \ref{setup_detail} is blocked, temporarily disabling the Franson interference. By performing a ToFS measurement of the single-mode spectrum of the $s'$-mode, we can directly observe the frequency distribution of the photon pair.

In Fig. \ref{photon_degenerate}, we conduct the aforementioned test using a demonstration PPLN-WG (not the crystal used in this study). In Fig. \ref{photon_degenerate}(a), the SPDC crystal is initially set to a non-degenerate state, where the frequency exchange mechanism causes the non-degenerate photon pairs to exhibit two dominant distributions in the $s'$-mode single-mode spectrum. As the crystal temperature varies, the photon pairs gradually approach the degenerate state, and HOM interference becomes increasingly pronounced. Ultimately, when the photon pairs reach degeneracy, the single-mode spectrum exhibits a single dominant distribution and reveals a comb-like structure, as shown in Figs. \ref{photon_degenerate}(b)-\ref{photon_degenerate}(c). Therefore, considering the method above allows us to fine-tune the temperature of the crystal, verify the degeneracy of photon pairs, and observe the transition of HOM interference. 

On the other hand, since this method requires tuning the photon wavelength, the fiber filter is removed in Fig. \ref{photon_degenerate}. As there is no filter applied to the spectrum of $s'$-mode photons, the spectral distribution retains a sinc function behavior, allowing the sidebands of the sinc function to be observed. It can be observed that the sidebands of the sinc function, due to their deviation from the main peak, influence the waveform of the single-mode spectrum in the non-degenerate case. In the degenerate case, these sidebands also appear on both sides of the main peak, slightly affecting the spectrum of the satellite state. Therefore, filtering out these sidebands using a fiber filter is a crucial aspect of this experiment.

\section{Controlling the phase of Franson interference}\label{phase_ctrl}

In Fig. \ref{TPI_7_11bin}(e) and \ref{TPI_7_11bin}(f), we implemented an alternative phase control scheme. For Fig. \ref{TPI_5bin}(b) and Fig. \ref{TPI_7_11bin}(b), the phase of the Franson interferometer is controlled by $\tau_F$, which is experimentally achieved by adjusting the voltage of the pizeo. However, in a typical TPI setup, the phase is jointly correlated with the biphoton frequencies, following the relation $\phi_F=\Omega_+\tau_F$ \cite{park2018time,ali2006experimental}. In an extreme case where the pump bandwidth is much narrower than the SPDC bandwidth, $\Omega_s$ and $\Omega_i$ will remain perfectly anti-correlated, allowing $\Omega_s+\Omega_i$ to be replaced by the pump frequency $\Omega_p$ \cite{ali2006experimental}. In our experiment, the linewidth of the pump laser is estimated to be a few kHz, which is significantly narrower than the bandwidth (estimated by the fitting method of $\Delta\omega_S\sim300$ GHz) of the SPDC crystal we used. This implies that we can further achieve phase control of the Franson interferometer by adjusting the pump wavelength.

To ensure that the operating wavelength of the pump laser remains both controllable and stable, in Fig. \ref{TPI_7_11bin}(e) and \ref{TPI_7_11bin}(f), the pizeo is fixed at a stable voltage, while the pump laser is continuously stabilized using feedback from a wavemeter. Next, the output wavelength is controlled by adjusting the wavelength locking position of the wavemeter, thereby modifying the interferometer phase.

Furthermore, although changing the pump frequency will slightly affect the PMC of the SPDC crystal, the large optical path difference between the short and long arms of the Franson interferometer significantly reduces the required variation in the pump laser frequency. In our system, the pump frequency is scanned over a range of approximately 149.8 MHz with a resolution of about 6.8 MHz, covering roughly two periods of the TPI. This range is much smaller than the SPDC bandwidth, ensuring that PMC drift can be considered negligible under our experimental conditions.

\section{TPI for each frequency bin in HFEQs}\label{SM_TPI_FB}
In this section, we present the TPI fringes for each frequency bin of the HFEQs, as shown in Fig. \ref{TPI_FB}. These high-contrast TPI interference patterns demonstrate that each discrete frequency region within the HFEQs retains the characteristic of CV frequency entanglement.

In Fig. \ref{TPI_FB}, we observe another intriguing phenomenon. The TPI patterns for all frequency bins within each HFEQ are in phase. This indicates that the frequency correlations exhibit a well-defined anti-correlated characteristic, corresponding to a -45-degree distribution in the joint spectrum. In contrast, if the JSI were not perfectly anti-correlated, its distribution in the joint spectrum would not be strictly aligned along the $\Omega_{-}$ direction. As a result, Franson interference along the $\Omega_{+}$ direction could introduce varying phase shifts in the TPI for different frequency bins, leading to an out-of-phase behavior in the TPI patterns. Our simulation support this statement.

\bibliography{refs}
\end{document}